\pdfoutput=1
\documentclass[10pt]{article}
\usepackage[T1]{fontenc}
\usepackage{amsmath,amssymb,graphicx}
\usepackage{hyperref}
\usepackage[font=small,labelfont=bf]{caption}
\usepackage{longtable}
\usepackage{verbatim}
\usepackage{amsfonts}
\usepackage{cite}

\usepackage[usenames,dvipsnames]{xcolor}
\definecolor{darkerblue}{rgb}{0.2,0.2,0.5}

\usepackage{afterpage}

\renewcommand{\arraystretch}{1.25}
\newcommand{\bear}{\begin{array}}
\newcommand{\ear}{\end{array}}
\newcommand{\be}{\begin{equation}}
\newcommand{\ee}{\end{equation}}
\newcommand{\beq}{\begin{eqnarray}}
\newcommand{\eeq}{\end{eqnarray}}
\newcommand{\beqa}{\begin{eqnarray}}
\newcommand{\eeqa}{\end{eqnarray}}

\def\OMIT#1{{}}
\newcommand{\lsim}{\mathrel{\rlap{\lower4pt\hbox{\hskip1pt$\sim$}}
    \raise1pt\hbox{$<$}}}         
\newcommand{\gsim}{\mathrel{\rlap{\lower4pt\hbox{\hskip1pt$\sim$}}
    \raise1pt\hbox{$>$}}}         
\def\simlt{\stackrel{<}{{}_\sim}}
\def\simgt{\stackrel{>}{{}_\sim}}
\newcommand{\tr}{{\rm Tr}}

\usepackage{tikz}
\usetikzlibrary{trees}
\usetikzlibrary{decorations.pathmorphing}
\usetikzlibrary{decorations.markings}
\tikzset{
    photon/.style={decorate, decoration={snake,amplitude=3pt,segment length=8pt}, draw=black},
    wino/.style={draw=redwine},    
    fermion/.style={draw=black, postaction={decorate},
        decoration={markings,mark=at position .55 with {\arrow[draw=black,scale=1,#1]{>}}}},
    scalar/.style={draw=black, dashed,postaction={decorate},
        decoration={markings,mark=at position .55 with {\arrow[draw=black,scale=1,#1]{>}}}},
    gluon/.style={decorate, draw=black,
        decoration={coil,amplitude=3pt, segment length=4pt}},
    graviton/.style={decorate, draw=black,
        decoration={zigzag,amplitude=3pt, segment length=4pt}}
}
\tikzstyle{blob}=[circle,
                              minimum size=0.01,
                              draw=black!80,
                              fill=black!80]   
\tikzstyle{redblob}=[circle,
                              thick,
                              minimum size=0.4cm,
                              draw=red!80,
                              fill=red!60] 

\usepackage{titlesec}

\title{\bf \Large SUSY Higgs Mass and Collider Signals with a Hidden Valley}

\author{Yuichiro Nakai$^a$, Matthew Reece$^a$, and Ryosuke Sato$^{b,c}$\\
{\small \color{gray} \texttt{ynakai, mreece @physics.harvard.edu, rsato@post.kek.jp}}\\
{$^a$ \em Department of Physics, Harvard University, Cambridge, MA 02138, USA}\\
{$^b$ \em Institute of Particle and Nuclear Studies, High Energy Accelerator}\\
{\em Research Organization (KEK), Tsukuba 305-0801, Japan}\\
{$^c$ \em Department of Particle Physics and Astrophysics,}\\
{\em Weizmann Institute of Science, Rehovot 76100, Israel}\vspace{0.3cm}}

\usepackage[utf8]{inputenc}
\begin{document}
\maketitle

\vspace{-9.5cm}
\begin{flushright}
{\footnotesize KEK-TH-1866} 
\end{flushright}
\vspace{8.1cm}

\begin{abstract}

\vspace{0.3cm}

We propose a framework of supersymmetric extensions of the Standard Model that can ameliorate both the SUSY Higgs mass problem and the missing superpartner problem.
New vectorlike matter fields couple to the Higgs and provide new loop contributions to its mass.
New Yukawa couplings are sizable and large supersymmetry breaking is not needed to lift the Higgs mass.
To avoid a Landau pole for the new Yukawa couplings, these fields are charged under a new gauge group, which confines and leads to a Hidden Valley-like phenomenology.
The Hidden Valley sector is almost supersymmetric and ordinary sparticles decay to exotic new states which decay back to Standard Model particles and gravitinos with reduced missing energy.
We construct a simplified model to simulate this scenario
and find a viable parameter space of specific benchmark models which ameliorates both of the major phenomenological problems with supersymmetry.

\end{abstract}

\section{Introduction}
\label{sec:intro}

The Standard Model Higgs boson, as a weakly interacting scalar particle, introduces a fine-tuning puzzle.
Supersymmetry remains an interesting possible resolution for this puzzle, despite increasingly strong constraints~\cite{Craig:2013cxa}.
Experimental results pose two significant obstacles to weak-scale supersymmetry as a solution to the fine-tuning problem.
The first is the {\em SUSY Higgs mass problem}. As is well-known, in the MSSM a 125 GeV Higgs mass requires large loop contributions from stops,
either from large $A$-terms or from very heavy unmixed stop masses (see~\cite{Haber:1990aw,Okada:1990vk,Barbieri:1990ja} for early references and~\cite{Draper:2011aa} and references therein for more recent work).
The second is the {\em missing superpartner problem}~\cite{Arvanitaki:2012ps}: experimental searches have so far failed to find a single superpartner.
(A small subset of the powerful recent searches for squarks and gluinos includes~\cite{Aad:2013wta,Aad:2014bva,Aad:2014kra,Chatrchyan:2014lfa,Khachatryan:2014doa,Khachatryan:2015wza}.)
A flurry of literature has attempted to solve both of these problems. The Higgs mass problem could be solved with new tree-level interactions beyond the MSSM \cite{Drees:1987tp,Drees:1988fc,Espinosa:1991gr,Randall:2002talk,Batra:2003nj,Casas:2003jx,Barbieri:2006bg,Dine:2007xi}.
It could also be solved by relaxing our fine-tuning requirements; generating sufficiently large $A$-terms in the MSSM is an interesting problem in its own right \cite{Evans:2011bea,Kang:2012ra,Craig:2012xp,Abdullah:2012tq,Evans:2013kxa,Basirnia:2015vga,Knapen:2015qba}. But another possibility is that stops are not the only important loop contributions:
vectorlike matter beyond the MSSM could help raise the Higgs mass, as studied in \cite{Moroi:1991mg,Moroi:1992zk, Babu:2004xg,Babu:2008ge,Martin:2009bg,Graham:2009gy,Martin:2010dc,
Martin:2010kk,Asano:2011zt,Endo:2011mc,Heckman:2011bb,Moroi:2011aa,Endo:2011xq,Endo:2012rd,Evans:2012uf,Martin:2012dg,Kitano:2012wv,Endo:2012cc,Lalak:2015xea}.
The missing superpartner problem, on the other hand, is generally addressed by modifying the dominant decays of superpartners.
$R$-parity violation \cite{Barbier:2004ez,Csaki:2011ge,Graham:2012th,Csaki:2013jza,Graham:2014vya,Heidenreich:2014jpa}, supersymmetric Hidden Valleys
\cite{Strassler:2006im,Strassler:2006qa}, Stealth Supersymmetry \cite{Fan:2011yu,Fan:2012jf,Stealth3}, compressed spectra \cite{Martin:2007gf,Baer:2007uz,Martin:2008aw,LeCompte:2011fh,Dreiner:2012gx,Dutta:2013gga,Bhattacherjee:2013wna},
supersoft supersymmetry \cite{Fox:2002bu,Nelson:2002ca,Kribs:2007ac,Kribs:2012gx,Alves:2015kia}, and theories with multiple invisible particles per decay chain \cite{Alves:2013wra} could all provide partial explanations for the absence of obvious signals in the data so far.

Most supersymmetric models beyond the MSSM that provide new interactions to lift the Higgs mass do not dramatically change the superpartner cross sections or decay chains in a way that can evade direct searches. Similarly, most models that alter superpartner decay chains to evade the missing superpartner problem do not involve interactions that lift the Higgs mass.
As a result, these two problems are usually treated as independent: attempts to construct natural SUSY models that agree with all existing data generally involve multiple modules that solve different problems.

In this paper we explore a scenario that can ameliorate both the SUSY Higgs mass problem and the missing superpartner problem. This way of lifting the Higgs mass has been previously studied by Babu, Gogoladze, and Kolda \cite{Babu:2004xg} and by Martin \cite{Martin:2010kk}, but it deserves renewed attention in the current phenomenological context in which we know the Higgs mass and that superpartner signals are absent so far. The idea is to add new vectorlike matter fields that couple to the Higgs and provide new loop contributions to its mass
as studied in refs.~\cite{Moroi:1991mg,Moroi:1992zk, Babu:2004xg,Babu:2008ge,Martin:2009bg,Graham:2009gy,Martin:2010dc,
Martin:2010kk,Asano:2011zt,Endo:2011mc,Heckman:2011bb,Moroi:2011aa,Endo:2011xq,Endo:2012rd,Evans:2012uf,Martin:2012dg,Kitano:2012wv,Endo:2012cc,Lalak:2015xea}.
The new matter fields are $\Psi_u$, ${\bar \Psi}_d$ in $SU(2)_L$ doublets and $\Psi$, ${\bar \Psi}$ in $SU(2)_L$ singlets.
Then they admit supersymmetric mass terms ($\mu$-terms), $W \supset m \Psi_u {\bar \Psi}_d + m' \Psi {\bar \Psi}$,
but (for appropriate hypercharge choices) they can also have Yukawa couplings to the Higgs fields: $W \supset \lambda_u H_u {\bar \Psi}_d \Psi + \lambda_d H_d \Psi_u {\bar \Psi}$.
If the Yukawa couplings $\lambda_{u,d}$ are fairly large, and also the supersymmetry breaking contributions to the masses of the new particles are of the same order as the supersymmetric masses $m, m'$,
then the threshold corrections to the Higgs quartic induced by integrating out these new particles can be significant.
All the new particles are near the weak scale, and they are interesting targets for collider searches.

The large Yukawa coupling $\lambda_{u,d}$ is crucial to lift up the Higgs boson mass.
For example, Martin pointed out that $\lambda_u \sim 1 $ can be obtained naturally
for several representation of $\Psi$'s under $SU(3)_C \times SU(2)_L \times U(1)_Y$ \cite{Martin:2009bg}.
However,
when the Yukawa couplings $\lambda_{u,d}$ are larger than that value,
we encounter one problem:
renormalization group running of these couplings hits a Landau pole immediately.
A possible solution to this problem is to introduce a new gauge interaction to the new particles
\cite{Babu:2004xg,Martin:2010kk,Heckman:2011bb,Evans:2012uf,Kitano:2012wv} (a variation with a new spontaneously broken gauge group was considered in \cite{Kyae:2013hda}).
As discussed in \cite{Martin:2010kk},
this gauge group confines and leads to a Hidden Valley-like phenomenology.
The author of \cite{Martin:2010kk} assumed that SUSY breaking in the Hidden Valley sector is of the same order with that in the ordinary sector.
Instead, we consider that the Hidden Valley sector is almost supersymmetric, as in Stealth Supersymmetry \cite{Fan:2011yu,Fan:2012jf},
which is naturally realized by viable mechanisms of SUSY breaking such as gauge mediation \cite{Giudice:1998bp} and gaugino mediation \cite{Kaplan:1999ac,Chacko:1999mi}.
Then, ordinary superpartners decay to exotic new states decaying back to Standard Model particles and gravitinos with reduced missing energy.
This is essential for hiding supersymmetry at the LHC.
We propose specific benchmark models and will publicly release tools based on a simplified model to simulate this scenario,
which will facilitate experimental searches.

The rest of the paper is organized as follows.
In section~\ref{sec:higgsmass}, we present our framework and analyze the effect on the Higgs mass in specific models.
We show a parameter space which can explain the correct Higgs mass without a Landau pole problem.
We assume a relatively low cutoff scale compared to the usual scale of the gauge coupling unification.
This can be justified by considering multi-fold replication of the SM gauge groups which realizes the accelerated unification and naturally leads to gaugino mediation as the SUSY-breaking inputs in the present framework.
From this background, we consider new vectorlike fields which respect the unification.
In section~\ref{sec:confinement}, Hidden Valley spectroscopy is firstly discussed.
We then present a simplified model for collider phenomenology which will be useful for later discussions.
We also analyze the effect on Higgs decays.
In section~\ref{sec:rge}, we use the SARAH code to give detailed numerical results for some benchmark models.
In section~\ref{sec:collider},  we demonstrate that the models hide from existing searches by showing some exclusion curves.
We also comment on phenomenology of the vectorlike fields.
In section~\ref{sec:conclude}, we conclude and comment on future directions
including the possibility of multi-fold replication of the SM gauge groups.

\section{Raising the Higgs mass}
\label{sec:higgsmass}

We here explain our framework with new loop contributions to the Higgs mass.
Then, we present specific models and calculate the Higgs mass from the Coleman-Weinberg potential.
We plot a parameter space which can explain the correct Higgs mass without a Landau pole problem.

\subsection{New loop contributions}

The minimal supersymmetric extension of the Standard Model generally predicts a light Higgs mass.
To obtain the observed 125~GeV Higgs mass in the MSSM,
we need a significant correction 
to the Higgs quartic coupling from top/stop loops,
\begin{equation}
\Delta \lambda_{H_u} \approx \frac{y_t^4 N_c}{16\pi^2}\left(\ln \frac{m_{\tilde{t}}}{m_t} + \frac{X_t^2}{m_{\tilde t}^2} - \frac{1}{12} \frac{X_t^4}{m_{\tilde t}^4}\right)\, ,  \label{Higgsquartictop} 
\end{equation}
where $N_c = 3$ and $X_t \equiv A_t - \mu \cos \beta$.
This contribution is sizable when the stop mass $m_{\tilde{t}}$ is large.
On the other hand, the large stop mass generates
a large quadratic term in the Higgs potential,
\begin{equation}
\Delta m_{H_u}^2 \approx -\frac{y_t^2 N_c}{8\pi^2} \left(m_{\tilde{t}_L}^2 
+m_{\tilde{t}_R}^2  + X_t^2\right)\ln \frac{M_{\rm m}^2}{m_{\tilde{t}}^2} \, , \label{Higgsquadratictop}
\end{equation}
where $M_{\rm m}$ is a scale at which the stop mass is generated.
This quadratic term has to be tuned away for the correct electroweak symmetry breaking, $-\mu^2 - m_{H_u}^2 \approx M_Z^2 /2$.
Generally, the tuning is worse than a percent even without a large logarithm.
However, the radiative corrections of \eqref{Higgsquartictop} and \eqref{Higgsquadratictop} imply a possible way to avoid this problem
\cite{Kitano:2012wv}.
If we could increase $y_t$, the quartic coefficient $\Delta \lambda_{H_u}$ increases as $y_t^4$
while the quadratic coefficient $\Delta m_{H_u}^2$ only increases as $y_t^2$ so that the required tuning is relaxed.
Then, if we have new Higgs interactions such as
\begin{equation}
\Delta W \, = \,  \lambda_{u} H_u \bar{\Psi}_d {\Psi} + \lambda_d H_d \Psi_u \bar{\Psi} + m \Psi_u \bar{\Psi}_{d} + m' \Psi \bar{\Psi} \, , 
\end{equation}
where $\Psi_u$, ${\bar \Psi}_d$ are in $ SU(2)_L$ doublets and $\Psi$, ${\bar \Psi}$ in $SU(2)_L$ singlets,
and assume $\lambda_u$ is larger than the top Yukawa,
we can lift up the Higgs mass without large soft masses of the new scalars, which reduces fine-tuning.

Without any other interactions,
running of the new large Yukawa coupling hits a Landau pole immediately.
A possible solution to this problem is to introduce a new gauge interaction to the new particles.
We can illustrate this by considering the case of the top Yukawa $y_t$
where the dominant contributions to the running are given by
\begin{equation}
\frac{d y_t}{d \ln \mu} \, \simeq \, \frac{y_t}{16\pi^2} \left( 6 y_t^\ast y_t - \frac{16}{3} g_3^2 \right) .
\end{equation}
The two terms in the right hand side of the equation have opposite signs.
Then, if the new gauge coupling is somewhat strong, the Landau pole problem can be avoided.
The new gauge field only couples to the new matter fields, $\Psi_{u, \,d}$, $\Psi$ and $\bar{\Psi}$.
Below the mass scale of these matter fields, the new gauge dynamics finally confines.
This is exactly the supersymmetric version of the Hidden Valley scenario
\cite{Strassler:2006im,Strassler:2006qa,Han:2007ae,Strassler:2008bv} (For a review, see Ref.~\cite{Zurek:2010xf}). A similar logic has been considered before \cite{Babu:2004xg,Martin:2010kk,Kyae:2013hda}.
The scenario is also similar to the idea of ``quirks'' \cite{Kang:2008ea,Burdman:2008ek,Kribs:2009fy,Harnik:2011mv}.
As we will see in later sections, it is remarkable that ordinary superpartners decay to exotic new states in the Hidden Valley sector
and the missing superpartner problem is ameliorated as in Stealth Supersymmetry \cite{Fan:2011yu,Fan:2012jf,Stealth3}.

\subsection{The model}

\renewcommand{\arraystretch}{1.3}
\begin{table}[t]
\begin{center}
\begin{tabular}{c|cccc|cc}
 & $SU(N)_{H}$ & $SU(3)_{C}$ & $SU(2)_{L}$ & $U(1)_Y$ & scalar name & fermion name
 \\
 \hline
  $\Psi_u$  &  $\mathbf  N$ & $\mathbf 1$ & $\mathbf  2$ & $1/2$ & $\phi_u$ & $\psi_u$ \\
  $\bar{\Psi}_d$ & $\mathbf{\bar{N}}$ & $\mathbf 1$ & $\mathbf  2$ & $-1/2$ & ${\bar \phi}_d$ & ${\bar \psi}_d$ \\
  $f$  &  $\mathbf  N$ & $\mathbf 3$ & $\mathbf  1$ & $-1/3$ & $\phi_f$ & $\psi_f$ \\
  $\bar{f}$ & $\mathbf{\bar{N}}$ & $\mathbf{\bar{3}}$ & $\mathbf  1$ & $1/3$ & ${\bar \phi}_f$ & ${\bar \psi}_f$ \\
  $\Psi_i$ &  $\mathbf  N$ & $\mathbf 1$ & $\mathbf  1$ & $0$ & $\phi_i$ & $\psi_i$ \\
  $\bar{\Psi}_i$ & $\mathbf{\bar{N}}$ & $\mathbf 1$ & $\mathbf  1$ & $0$ & ${\bar \phi}_i$ & ${\bar \psi}_i$ \\
\end{tabular}
\end{center}
\caption{The charge assignments. For convenience we also list the names we use to refer to the (scalar and left-handed Weyl fermion) components of each chiral multiplet.
Notice, in particular, that we always use daggers for complex conjugation, whereas bars are simply part of the name of the field.}
\label{tab:model}
\end{table}
\renewcommand{\arraystretch}{1}

We consider a supersymmetric $SU(N)_{H}$ gauge theory with $5+F$ flavors; $\Psi_{u}$, $\bar{\Psi}_{d}$, $\Psi_i$, $\bar{\Psi}_i$, $f$ and $\bar{f}$
(the lower index $i = 0, 1, \ldots F-1$ is the flavor index---enumerated from zero for reasons that will become clear shortly).
The charge assignment is summarized in Table~\ref{tab:model}.
The colored vectorlike particles $f$, $\bar{f}$ are introduced to complete the $SU(5)$ multiplets.
The new superpotential terms involving the vectorlike particles beyond the MSSM are
\begin{equation}
W_{\rm VL} \, = \,  \lambda_{u,i} H_u \bar{\Psi}_d {\Psi}_{i} + \lambda_{d,i} H_d \Psi_u \bar{\Psi}_{i} + m \Psi_u \bar{\Psi}_{d} + m'_{ij} \Psi_i \bar{\Psi}_j  + M f \bar{f} \, .
\label{superpo} 
\end{equation}
We also have new soft supersymmetry breaking terms for the new particles:
\beq
-{\cal L}_{\rm soft,VL} & = & {\tilde m}_u^2 \phi_u^\dagger \phi_u + {\tilde m}_d^2 {\bar \phi}_d^\dagger {\bar \phi}_d  + {\tilde m}_{ij}^2 \phi_i^\dagger \phi_j + {\tilde {\bar m}}_{ij}^2 {\bar \phi}_i^\dagger {\bar \phi}_j + {\tilde m}_f^2 \phi_f^\dagger \phi_f
+ {\tilde {\bar m}}_f^2 {\bar \phi}_f^\dagger {\bar \phi}_f \nonumber \\
& + & \left(A_{u,i} H_u {\bar \phi}_d \phi_i + A_{d,i} H_d \phi_u {\bar \phi}_i + b_m \phi_u {\bar \phi}_d + b'_{m,ij} \phi_i {\bar \phi}_j + b_M \phi_f {\bar \phi}_f + {\rm h.c.}\right).
\eeq
This introduces a very large number of new parameters, so we will make some simplifying assumptions. We have a global symmetry $SU(F)_1 \times SU(F)_2$ under which $\Psi_i$ and ${\bar \Psi}_i$ transform as $(F,1)$ and $(1,{\bar F})$ respectively. 

The couplings to the Higgs bosons inevitably break the symmetry: the Yukawa couplings in $W_{\rm VL}$ transform as spurions in the $({\bar F},1)$ and $(1,F)$ representations of the full $SU(F)_1 \times SU(F)_2$. We will take the $A$-terms
and the Yukawa couplings to be aligned, which could be justified by gauge or gaugino mediation models. Then a global $SU(F-1)_1 \times SU(F-1)_2$ symmetry is preserved by these terms:
\beq
\lambda_{u,i} = \lambda_u \delta_{i0},~~\lambda_{d,i} = \lambda_d \delta_{i0},~~A_{u,i} = A_u \delta_{i0},~~A_{d,i} = A_d \delta_{i0}. 
\label{eq:Yukawaansatz}
\eeq
In other words, we will assume that $\Psi_0$ and ${\bar \Psi}_0$ are the only Standard Model singlets charged under $SU(N)_H$ that couple to the Higgs fields. (In the presence of the $\lambda$'s alone,
this would simply be a choice of label; in the presence of both $\lambda$'s and $A$'s, it is an assumption about the physics of SUSY breaking---made purely for simplicity.)

Now that we have already singled out $\Psi_0$ and ${\bar \Psi}_0$ as different from the other Standard Model singlets, let us consider an ansatz for the mass terms that is as simple as possible while still allowing these two fields to play a possibly different role than the others:
\beq
{\rm If}~i \neq 0~{\rm or}~j \neq 0:~~m'_{ij} = m' \delta_{ij},~~b'_{m,ij} = b'_m \delta_{ij},~~{\tilde m}_{ij}^2 = {\tilde m}^2 \delta_{ij},~~{\tilde {\bar m}}^2_{ij} = {\tilde {\bar m}}^2 \delta _{ij}. \nonumber \\
m'_{00} \equiv m'_0,~~~b'_{m,00} \equiv b'_{m,0},~~~{\tilde m}_{00}^2 \equiv {\tilde m}_0^2,~~~{\tilde {\bar m}}^2_{00} \equiv {\tilde {\bar m}}_0^2.
\label{eq:massansatz1}
\eeq
This ansatz, taken together with (\ref{eq:Yukawaansatz}), implies an unbroken $SU(F-1)$ diagonal symmetry. It assumes that the two fields $\Psi_0$ and ${\bar \Psi}_0$ which couple to the Higgs pair up through a vectorlike mass.
This assumption is made largely to avoid becoming burdened with too many arbitrary choices of parameters to consider, although one could try to justify it in a UV completion.
The fields $\Psi_i$ and ${\bar \Psi}_i$ with $i = 1,\ldots F-1$ transform in the fundamental and antifundamental, respectively, of the unbroken $SU(F-1)$ symmetry.
This ansatz for the soft masses (including $b$-terms) will arise in gauge or gaugino mediation, which guarantees universality of (for instance) the ${\tilde m}_{ij}^2$ terms,
while the additional Yukawa couplings of $\Psi_0$ and ${\bar \Psi}_0$ can split ${\tilde m}_0^2$ from ${\tilde m}^2$ in the running.
Alternatively, we could choose the special case $m'_0 = m', b'_{m,0} = b'_m, {\tilde m}_0^2 = {\tilde m}^2, {\tilde {\bar m}}_0^2 = {\tilde {\bar m}}^2$,
in which case the mass terms alone leave an $SU(F)$ diagonal symmetry unbroken and only the couplings to the Higgs further break it to $SU(F-1)$.

Because our ansatz leaves a large symmetry group unbroken, it can lead to unwanted stable particles in the theory.
In practice, then, we will use this ansatz as a simplifying assumption in discussing the Coleman-Weinberg potential and the renormalization group equations.
For phenomenological purposes, we will assume small variations in the mass terms for different flavors that break all remaining flavor symmetries and allow all particles in the fundamental representation of $SU(N)_H$ to eventually decay to the lightest such particle. Such an approximate symmetry that is slightly broken could be produced in a variety of UV completions.

\subsection{The Coleman-Weinberg potential}

\begin{figure}[!t]
  \begin{center}
          \includegraphics[width=1.0\textwidth]{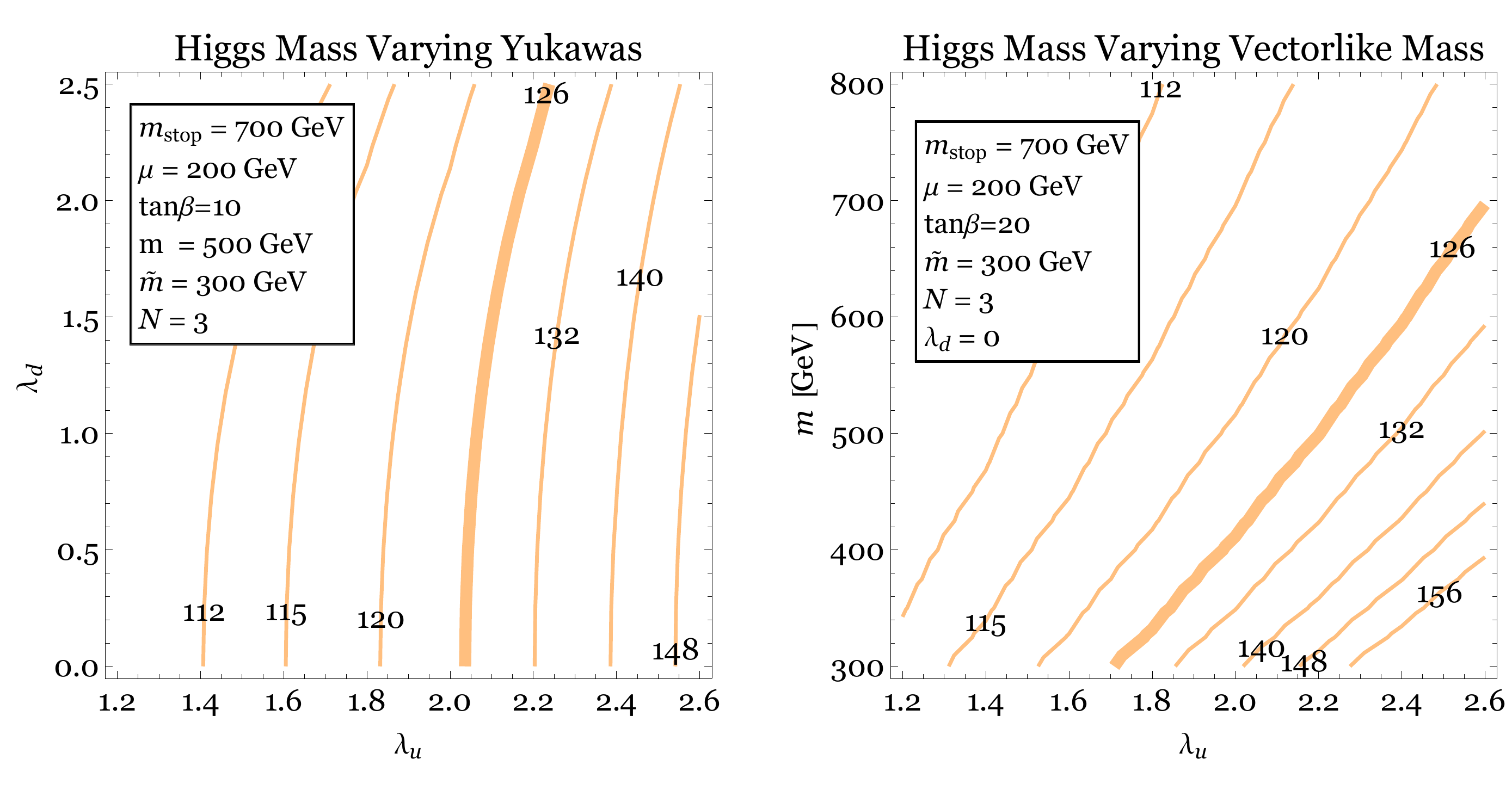}
    \caption{The effect of hidden sector vectorlike matter on the physical Higgs boson mass. We plot contours of constant physical Higgs mass as a function of various parameters of the Hidden Valley sector,
    with the simplifications $m_0' = m$ and ${\tilde m}_u^2 = {\tilde m}_d^2 = {\tilde {\bar m}}_0^2 = {\tilde m}_0^2 = {\tilde m}^2$ (with zero $A$ and $b$ terms).
    In the plot at left, we fix all masses and vary the two Yukawa couplings $\lambda_u$ and $\lambda_d$. We see that, due to large $\tan \beta$, the result is mostly insensitive to $\lambda_d$ unless it is very large.
    At right, we vary the supersymmetric vectorlike mass parameter $m$ (fixing the SUSY breaking mass ${\tilde m}$) as well as $\lambda_u$. We have set $\lambda_d = 0$ in this plot for simplicity, but a nonzero $\lambda_d$ will play an important role later in the paper.}
    \label{fig:liftinghiggsmass}
  \end{center}
\end{figure}

\begin{figure}[!t]
  \begin{center}
          \includegraphics[width=1.0\textwidth]{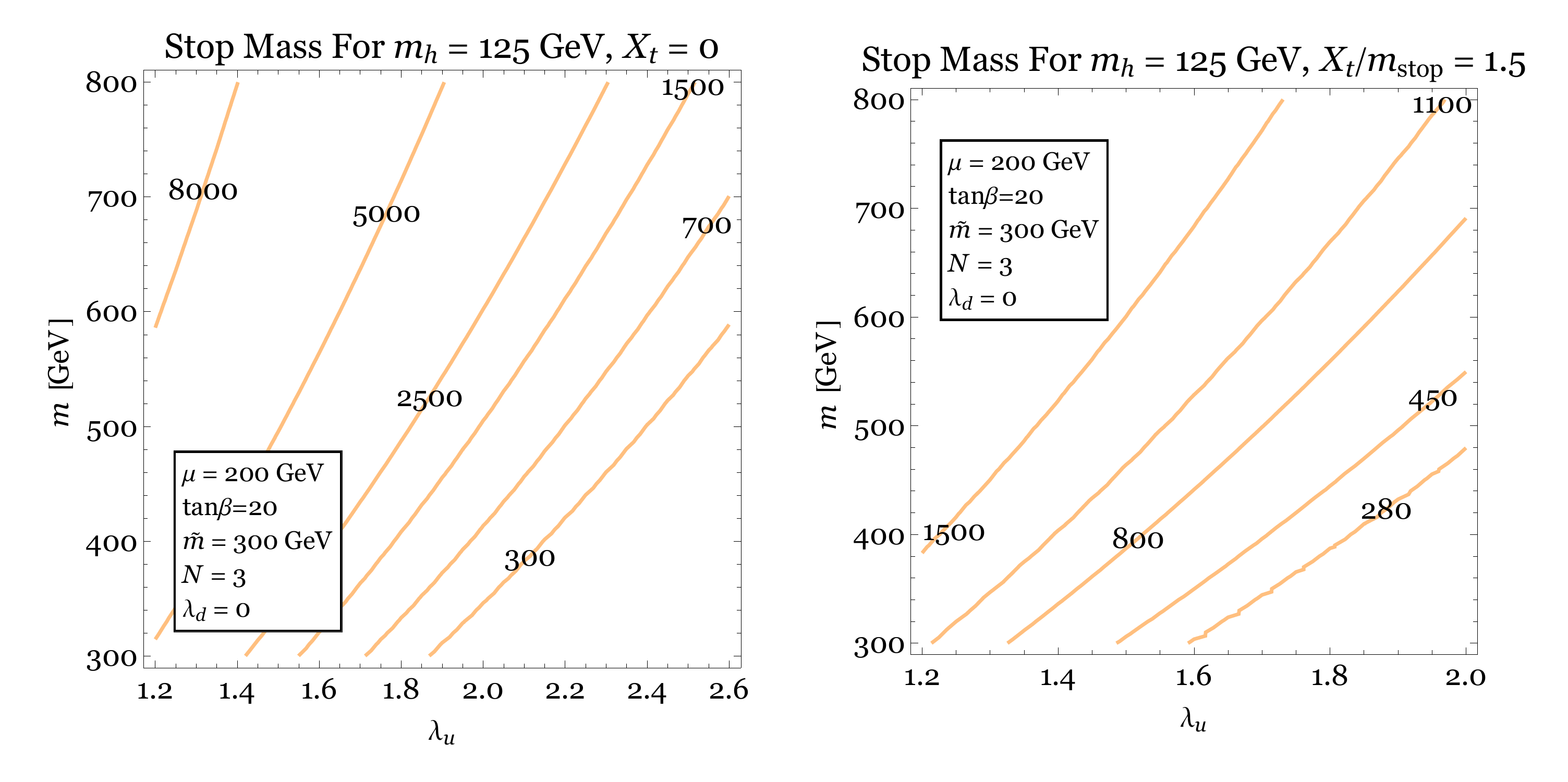}
    \caption{The stop mass scale $m_{\rm stop}$ (taken to be the geometric mean of the stop masses) necessary to raise the Higgs mass to 125 GeV,
    assuming that a contribution to the quartic coupling arises from a Hidden Valley sector with the specified parameters. At left, we assume no stop mixing ($X_t = 0$),
    so when the Hidden Valley contribution is turned off the stop masses must be several TeV. Vectorlike matter at a few hundred GeV with $\lambda_u \simgt 2$ can lift the Higgs to 125 GeV even with unmixed stops at 300 GeV.
    At right, we show a scenario with significant stop mixing $X_t = 1.5 m_{\rm stop}$, in which case the necessary values of $\lambda_u$ are smaller. (In particular, notice that the horizontal axis in the plot at right has a much smaller range!)
    Again, $\lambda_d$ was set to zero to give a simple illustration, but will be nonzero in the remainder of the paper.}
    \label{fig:neededstopmass}
  \end{center}
\end{figure}

We compute the one-loop effect on the Higgs potential using the general Coleman-Weinberg result,
\beq
V_{\rm CW} = \frac{1}{32 \pi^2} \tr \sum (-1)^F {\cal M}^4 \left(\log\frac{{\cal M}^2}{\mu^2} - \frac{3}{2}\right),
\eeq
with ${\cal M}$ the mass matrix of the new particles. The new contribution to the Higgs mass will be proportional to SUSY-breaking soft terms, ${\tilde m}_0^2, {\tilde {\bar m}}_0^2, {\tilde m}_u^2, {\tilde m}_d^2, A_u, A_d, b_m$, and $b'_{m,0}$.
It will also depend on the supersymmetric parameters $\lambda_u, \lambda_d, m, m'_0$, and $\mu$ (the Higgs superpotential mass).
The scalar mass matrix in the $(\phi_0, {\bar \phi_0}^\dagger, \phi_u, {\bar \phi}_d^\dagger)$ basis is:
\beq
{\cal M}_{\rm s}^2 = \begin{pmatrix}  {\tilde m}_0^2 + |m'_0|^2 + |\lambda_u H_u|^2 & b^{\prime \dagger}_{m,0} & \lambda_u^\dagger H_u^\dagger m + \lambda_d H_d m^{\prime \dagger}_0 & A_u^\dagger H_u^\dagger + \lambda_u^\dagger \mu H_d \\
b'_{m,0} & {\tilde {\bar m}}^2_0 + |m'_0|^2 + |\lambda_d H_d|^2 & A_d H_d + \lambda_d \mu^\dagger H_u^\dagger & \lambda_u^\dagger H_u^\dagger m'_0 + \lambda_d H_d m^\dagger \\
\lambda_u H_u m^\dagger + \lambda_d^\dagger H_d^\dagger m'_0 & A_d^\dagger H_d^\dagger + \lambda_d^\dagger \mu H_u & {\tilde m}_u^2 + |m|^2 + |\lambda_d H_d|^2 & b_m^\dagger \\
A_u H_u + \lambda_u \mu^\dagger H_d^\dagger& \lambda_u H_u m^{\prime \dagger}_0 + \lambda_d^\dagger H_d^\dagger m & b_m & {\tilde m}_d^2 + |m|^2 + |\lambda_u H_u|^2
\end{pmatrix}.
\eeq
The fermion mass matrix ${\cal M}^2_{\rm f}$ can be obtained by setting all of the SUSY-breaking parameters in ${\cal M}^2_{\rm s}$ to zero.
Due to the large number of free parameters, we will not attempt to give a complete analytic expression for the shift in the Higgs mass.
Here, we will present analytic answers for some special simplified ans\"atze for the couplings, and also some plots to illustrate the result.

In the limit that $A$- and $b$-terms are zero, $\mu$ is neglected, all soft masses are ${\tilde m}^2$, $m'_0 = m$, and $m^2 \gg {\tilde m}^2$, a simple computation based on the 1-loop effective K\"ahler potential~\cite{Brignole:2000kg} gives
\beq
V_{\rm quartic} \approx \frac{N {\tilde m}^2}{48 \pi^2 m^2} \left(|\lambda_u|^2 H_u^\dagger H_u - |\lambda_d|^2 H_d^\dagger H_d\right)^2. \label{eq:simplestquarticestimate}
\eeq
More general expressions can be derived, but are not very enlightening; for instance, retaining the effective K\"ahler potential approximation (i.e.~small SUSY breaking) but allowing the various masses to differ, one finds that the up-type Higgs quartic coupling is
\beq
V_{\rm quartic} & \approx & \frac{N \left|\lambda_u H_u\right|^4}{32 \pi^2} \left[\frac{|m'_0|^4({\tilde {\bar m}}^2_0 - 2 {\tilde m}_d^2 - {\tilde m}_0^2)
+ 5 |m|^2 |m'_0|^2 ({\tilde m}_0^2 + {\tilde {\bar m}}_0^2 - {\tilde m}_d^2 - {\tilde m}_u^2) + |m|^4 (2 {\tilde m}_0^2 + {\tilde m}_d^2 - {\tilde m}_u^2)}{(|m|^2-|m'_0|^2)^3}\right. \nonumber \\
& & ~~~~~+ \left. 2 {\tilde m}_0^2 \left(\frac{|m|^2 + |m'_0|^2}{(|m|^2 - |m'_0|^2)^2} + \frac{2 |m|^2 |m'_0|^2}{\left||m|^2 - |m'_0|^2\right|^3} \log\frac{|m|^2 + |m'_0|^2 - \left||m|^2 - |m'_0|^2\right|}{|m|^2 + |m'_0|^2 + \left||m|^2 - |m'_0|^2\right|}\right)\right].
\eeq
These analytic approximations are mostly useful to highlight some important qualitative points.
The expressions scale as SUSY-violating mass squared terms divided by SUSY-preserving mass squared terms.
Thus the SUSY-breaking splittings must not be too small, in order to produce a large effect. Recall that, if the light Higgs is mostly $H_u$, the measured Higgs mass requires a quartic coupling $V \supset \lambda |H|^4$ with $\lambda \approx 0.13$.
The equation (\ref{eq:simplestquarticestimate}) gives a contribution of $\approx 0.05$ to $\lambda$ when $N = 3, \lambda_u \approx 2$, and ${\tilde m}^2 \approx \frac{1}{2} m^2$.
This shows that achieving a sizable effect on the Higgs properties will require a large Yukawa coupling $\lambda_u$ in the superpotential.

Let us now be more quantitative. We have plotted the shift in the Higgs mass in Fig.~\ref{fig:liftinghiggsmass}.
For a different perspective on the same result, in Fig.~\ref{fig:neededstopmass} we show the value of the stop mass needed to lift the Higgs mass to 125 GeV for fixed parameters of the new vectorlike matter.
To compute the Higgs mass, we match the quartic coupling $\lambda$ of the Standard Model Higgs boson to the SUSY prediction,
including stop threshold corrections and the correction from the Coleman-Weinberg potential involving the new vectorlike matter fields, at an RG scale $\mu_{\rm R} = M_{\rm SUSY} \equiv \sqrt{m_{{\tilde t}_1} m_{{\tilde t}_2}}$.
Then we solve the one-loop Standard Model RG equation to run down to the electroweak scale and find a physical Higgs mass. This approximation resums the leading large logarithmic terms that are important at large stop mass.
We have greatly simplified the spectrum in the plot by fixing all soft masses to be equal at $M_{\rm SUSY}$, which gives a good guide for the qualitative size of the effect.
In later sections we will run RGEs from a higher scale and use the complete Coleman-Weinberg potential.
Finally let us comment on higher-order corrections on the Higgs boson mass.
Although they could be calculated thanks to recent development of \texttt{SARAH} code \cite{Goodsell:2014bna, Goodsell:2015ira, Nickel:2015dna},
for the discussion of phenomenology such as collider signals, the most important uncertainty comes from uncertainty of confinement scale.
Thus, in this paper, we simply take one-loop correction and neglect this uncertainty.

\subsection{Avoiding Landau poles}

\begin{figure}[!t]
\begin{center}
\includegraphics[clip, width=7cm]{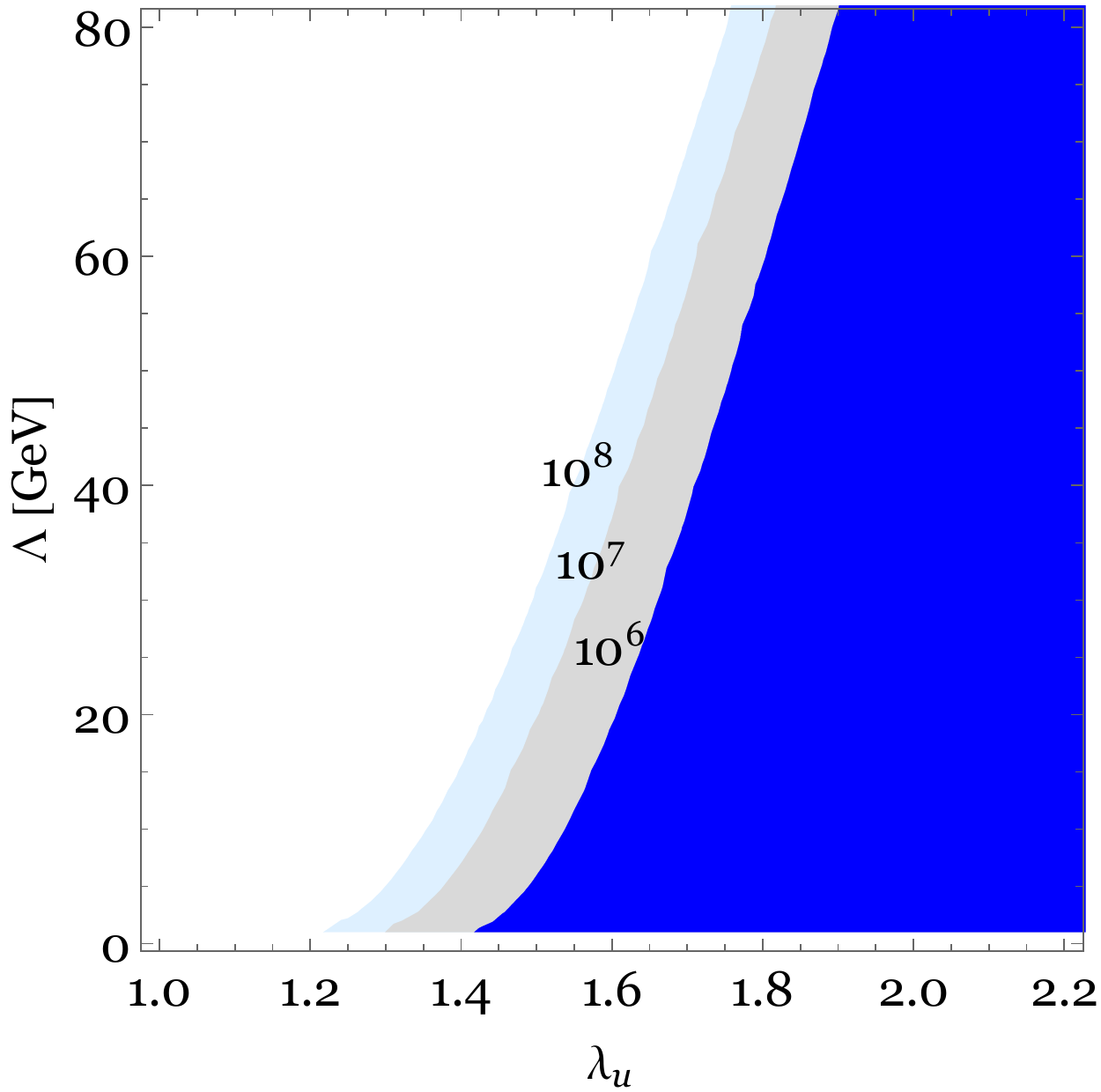} \hspace{1cm}
\includegraphics[clip, width=7cm]{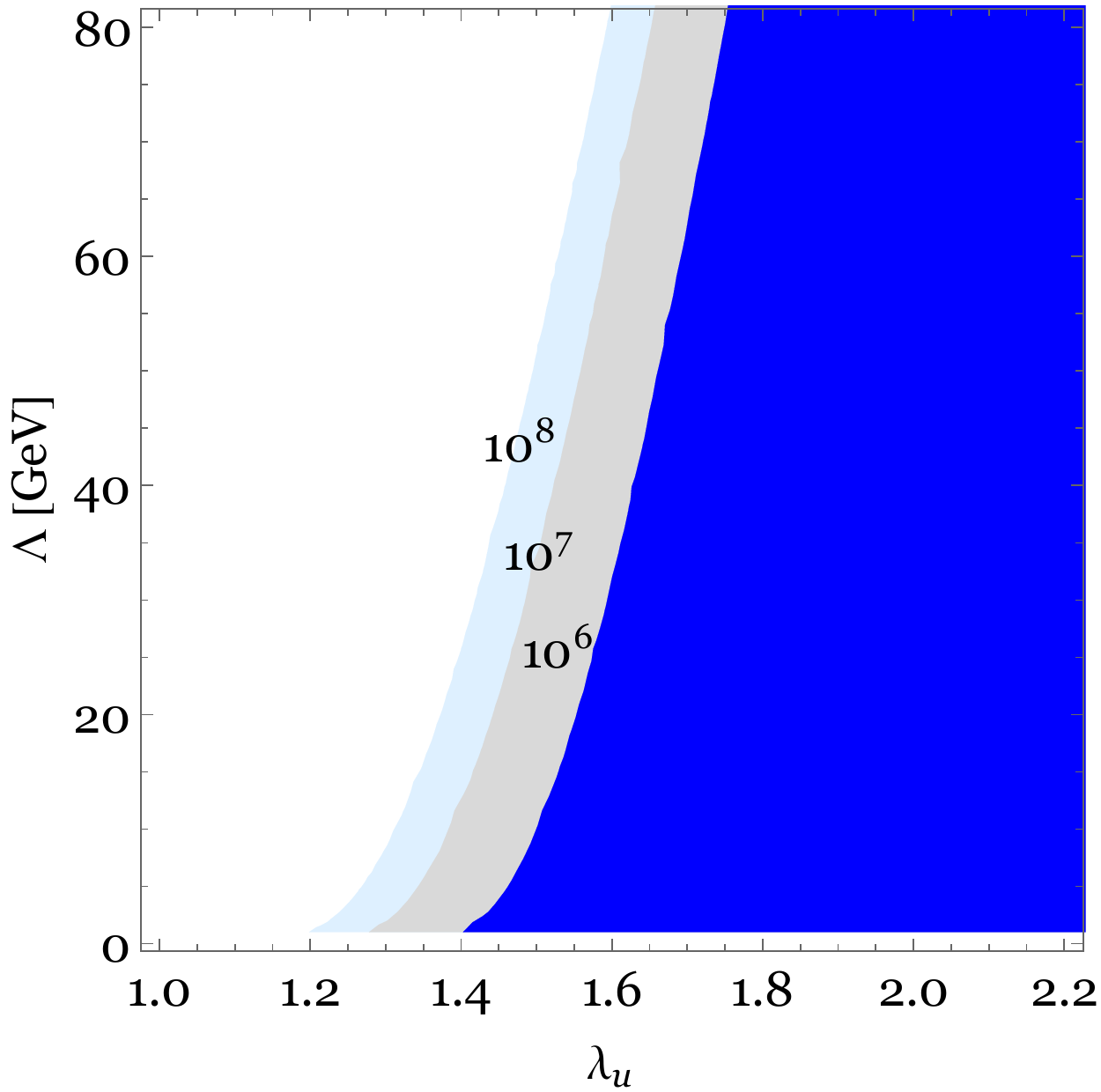} \vspace{0cm}
\vspace{0.5cm}
\includegraphics[clip, width=7cm]{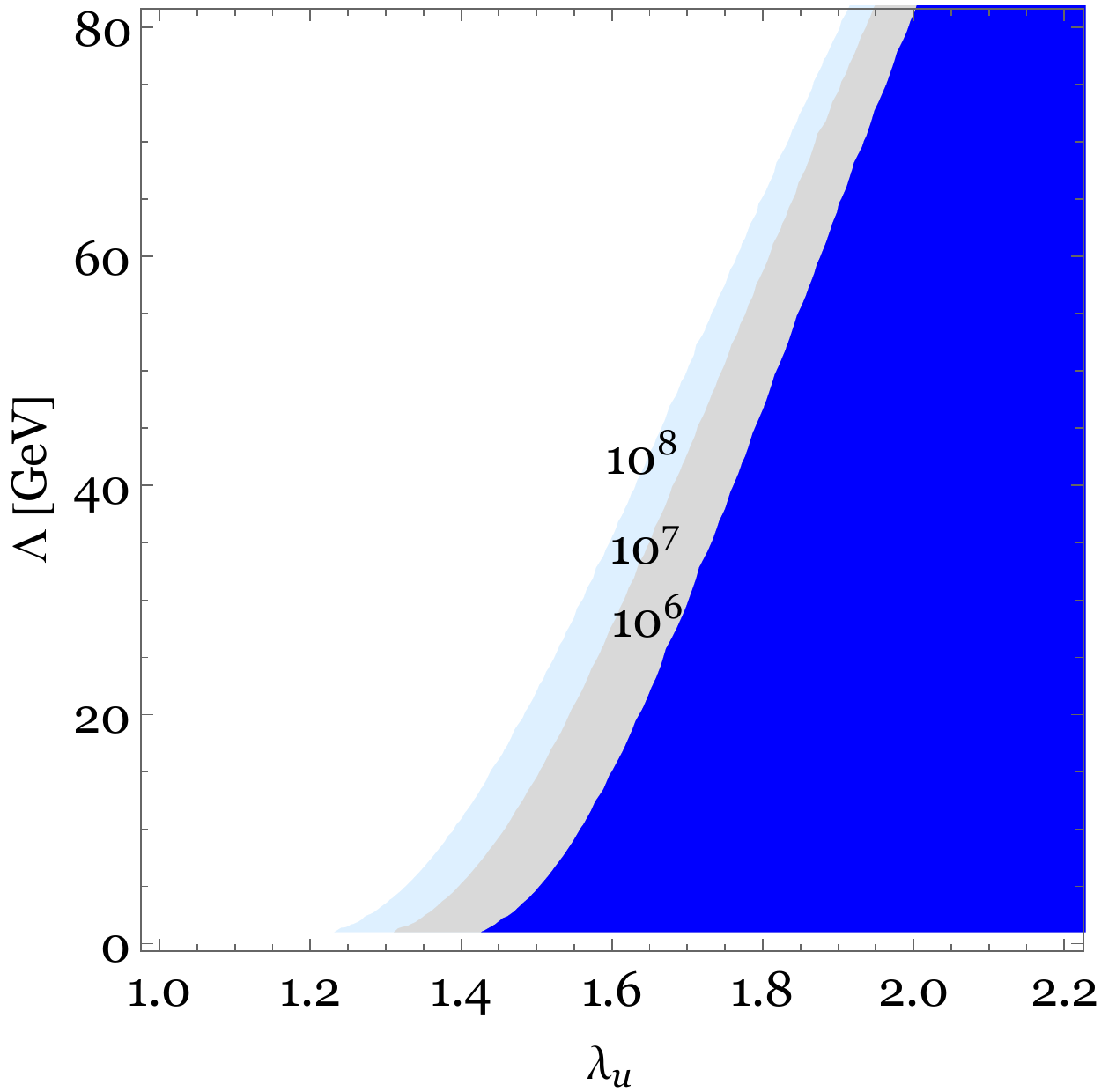} \hspace{1cm}
\includegraphics[clip, width=7cm]{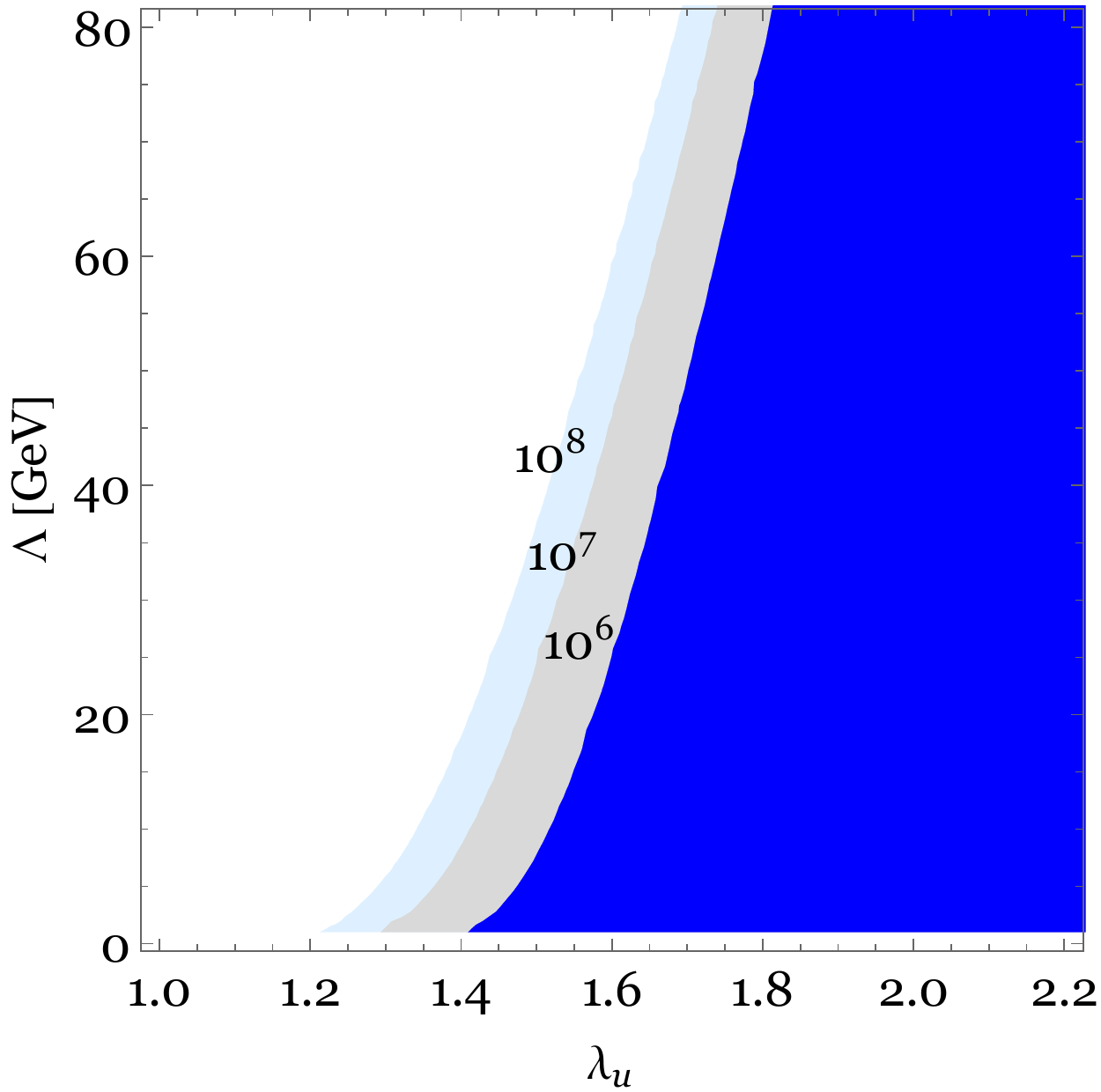} \vspace{0cm}
\caption{
The allowed (white) region of the coupling $\lambda_u$ at the vectorlike mass scale and the confinement scale $\Lambda$
where the Yukawa coupling $\lambda_u$ and the gauge coupling $g_h$ do not hit a Landau pole until $10^8$, $10^7$, $10^6 \, \rm GeV$ (from left to right).
The supersymmetric masses of the vectorlike fields are taken to be $m'_0 = m' = m = M = 300 \, \rm GeV$ in the left panel while $m'_0 = m' = m = M = 500 \, \rm GeV$ in the right panel.
The Yukawa coupling $\lambda_d$ at the vectorlike mass scale is assumed to be zero.
The number of the SM singlets is $F=2$ in the upper two panels and $F=3$ in the lower two panels.}
\label{fig:su3_3_result}
\end{center}
\end{figure}

A new gauge interaction keeps the
running of the new Yukawa couplings from hitting a Landau pole.
The new gauge dynamics finally confines.
We investigate a viable region of the confinement scale $\Lambda$ and the Yukawa coupling $\lambda_u$ where
the Landau pole problem is avoided.
For numerical analyses, we consider the 2-loop RGEs by using the SARAH code \cite{Staub:2008uz, Staub:2013tta}.
We ignore SUSY breaking and assume $m'_0 = m' = m = M$ for simplicity of the analyses.
After integrating out the vectorlike fields, the effective theory is pure SUSY Yang-Mills
and the (canonical, rather than holomorphic) gauge coupling below the vectorlike mass scale at two-loop level is given by
\begin{equation} \label{eq:runningalphah}
\begin{split}
\alpha_h (\mu) \equiv \frac{g_h^2 (\mu)}{4 \pi} &\approx \frac{4 \pi}{b_0 \ln ( {\mu^2}/{\Lambda^2} )} \Biggl( 1 - \frac{2 b_1}{b_0^2} \frac{\ln \left( \ln ( {\mu^2}/{\Lambda^2} ) \right)}{\ln ( {\mu^2}/{\Lambda^2} )} \\
&\qquad+ \frac{4 b_1^2}{b_0^4 \ln^2 ( {\mu^2}/{\Lambda^2} )} \left( \left( \ln \left( \ln ( {\mu^2}/{\Lambda^2} ) \right) - \frac{1}{2} \right)^2 - \frac{5}{4} \right) \Biggr), \\
\end{split}
\end{equation}
where $b_0 = 3 N$ and $b_1 = 3 N^2$ \cite{Chetyrkin:1997sg}.
Note that this expression is renormalization scheme independent and 
enables us to know the value of the gauge coupling at the vectorlike mass scale.
Then, we can judge if the theory with the vectorlike fields hits a Landau pole or not until some UV scale when we fix a confinement scale (and a Yukawa coupling $\lambda_u$ at the vectorlike mass scale).

Figure~\ref{fig:su3_3_result} shows the allowed (white) region of the coupling $\lambda_u$ at the vectorlike mass scale and the confinement scale $\Lambda$
where the Yukawa coupling $\lambda_u$ and the gauge coupling $g_h$ do not hit a Landau pole until $10^8$, $10^7$, $10^6 \, \rm GeV$ (from left to right).
The number of SM singlets is $F=2$ in the upper two panels and $F=3$ in the lower two panels.
The supersymmetric masses of the vectorlike fields are taken to be $m'_0 = m' = m = M = 300 \, \rm GeV$ in the left panels and $m'_0 = m' = m = M = 500 \, \rm GeV$ in the right panels.
The Yukawa coupling $\lambda_d$ at the vectorlike mass scale is assumed to be zero.
From these figures, we can know a lower bound of the confinement scale $\Lambda$ for a fixed value of the Yukawa coupling $\lambda_u$.
The bound is weaker as we lower the UV cutoff scale.
Smaller supersymmetric masses of the vectorlike fields also contribute to lowering the bound of the confinement scale.
The lower bound for $F=3$ is weaker than that for $F=2$.
To obtain the correct Higgs mass, a relatively large Yukawa coupling is required, $\lambda_u \gsim 1.5$.
In the lower right panel of Figure~\ref{fig:su3_3_result}, for example, we can see that this is realized when the confinement scale is $\Lambda\gsim 10 \, \rm GeV$ for a cutoff scale $10^6 \, \rm GeV$.
We will use these results in later sections.

We have assumed a relatively low cutoff scale $10^{6, 7, 8} \, \rm GeV$ compared to the usual scale of the gauge coupling unification around $10^{16} \, \rm GeV$.
This can be justified by considering multi-fold replication of the SM gauge groups which naturally leads to gaugino mediation as the SUSY-breaking inputs in the present framework.
We will further comment on this possibility in the final section although the detailed analyses are left for future work.

\section{Phenomenology with the Hidden Valley}
\label{sec:confinement}

In this section, spectroscopy of the Hidden Valley in our scenario is discussed.
We try to specify the glueball and gluinoball spectra. 
We then present a simplified model for collider phenomenology with the Hidden Valley which will be useful for later discussions.
We analyze how the lightest neutralino and the Hidden Valley particles decay by using the presented simplified model.
We also investigate the Higgs boson decays.

\subsection{Phenomenological possibilities}

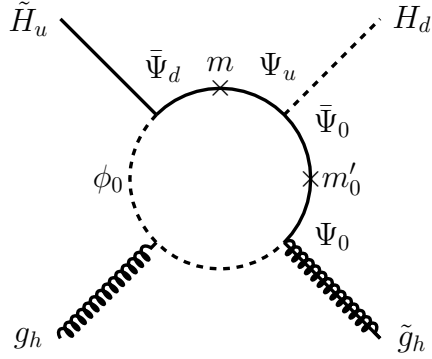
\begin{figure}[!h]
\begin{center}
\scalebox{0.85}{
\begin{tikzpicture}[line width = 1.5 pt]
\node at (-3.0,2.5) {\Large ${\tilde H}_u$};
\draw (-2.5,2.5)--(-1,1);
\draw[gluon] (-2.5,-2.5)--(-1,-1);
\node at (-3.0,-2.5) {\Large $g_h$};
\draw[gluon] (1,-1)--(2.5,-2.5);
\draw (1,-1)--(2.5,-2.5);
\node at (3.0,-2.5) {\Large ${\tilde g}_h$};
\draw[dashed] (1,1)--(2.5,2.5);
\node at (3.0,2.5) {\Large $H_d$};
\draw [domain=-45:135] plot ({1.414*cos(\x)},{1.414*sin(\x)});
\draw [dashed, domain=135:315] plot ({1.414*cos(\x)},{1.414*sin(\x)});
\node at (-1.75,0) {\Large $\phi_0$};
\node at (1.414,0) {\Large $\times$};
\node at (1.9,0) {\Large $m'_0$};
\node at (1.75,0.9) {\Large ${\bar \Psi}_0$};
\node at (1.75,-0.9) {\Large $\Psi_0$};
\node at (0,1.414) {\Large $\times$};
\node at (0,1.8) {\Large $m$};
\node at (-0.9,1.75) {\Large ${\bar \Psi}_d$};
\node at (0.9,1.75) {\Large $\Psi_u$};
\end{tikzpicture}
}
\end{center}
\caption{One of the microscopic interactions responsible for decays of $R$-parity odd particles into the hidden sector. Replacing the Higgs boson by its VEV, this becomes a decay of the neutral higgsino to a hidden-sector gluon and gluino.}
\label{fig:loopdiagram}
\end{figure}

\begin{figure}[!h]
\begin{center}
\begin{tikzpicture}[line width = 1.5 pt]
\node at (-1.0,-0.5) {\Large ${\tilde \chi}^0$};
\draw (-2.0,0.0)--(0.0,0.0);
\draw[dashed] (-4.0,0.0)--(-2.0,0.0);
\node at (-3.0,-0.5) {\Large ${\tilde q}$};
\draw (-2.0,0.0)--(-1.0,1.0); 
\node at (-0.9,1.5) {\Large $q$};
\draw[dashed] (0,0)--(1,1); \node at (0.4,0.8) {\Large $S$};
\draw[fermion] (1,1)--(2,2); \node at (2,2.5) {\Large $b$};
\draw[fermion] (2.414,1)--(1,1); \node at (2.4,1.5) {\Large $\bar b$};
\draw (0,0)--(2.014,0);
\node at (1.0,-0.5) {\Large ${\tilde S}$};
\draw[dashed] (2.014,0)--(3.239,0.707);
\node at (2.3,0.5) {\Large $S$};
\draw (2.014,0)--(3.428,0);
\node at (2.7,-0.5) {\Large ${\tilde G}$};
\draw[fermion] (3.239,0.707)--(4.239,1.707); \node at (4.2,2.2) {\Large $b$};
\draw[fermion] (4.653,0.707)--(3.239,0.707); \node at (5.0,0.707) {\Large $\bar b$};
\end{tikzpicture}
\end{center}
\caption{A squark decay process in the scenario where the confinement scale is sufficiently large that we should think of the decay as directly into composite states. In this case, we obtain a decay ${\tilde \chi}^0 \to b{\bar b}b{\bar b}{\tilde G}$, again with a soft gravitino through the Stealth SUSY mechanism and with pairs of $b$-jets reconstructing the scalar $S$ or $S'$. This scenario is the focus of our work in this paper.}
\label{fig:nopartonshower}
\end{figure}
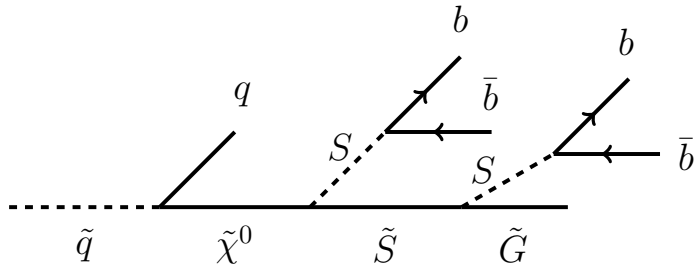

The basic scenario we have described will always raise the Higgs boson mass and lead to cascade decays of supersymmetric particles into the dark sector, proceeding through one-loop interactions like the one depicted in figure \ref{fig:loopdiagram}. However, the precise details of the phenomenology depend on several choices we can make in constructing the scenario, including:
\begin{itemize}
\item {\em Stealth SUSY or not.} If SUSY breaking is mediated only weakly to states with no Standard Model charges---as in gauge or gaugino mediation via messengers that have SM charges but no $SU(N)_h$ charge---the hidden sector can be nearly supersymmetric. This gives a realization of Stealth Supersymmetry. Even if SUSY is mediated directly to the hidden sector, bounds on superpartners could still be weaker as the missing energy can be diluted by the number of particles, as in Hidden Valley scenarios \cite{Strassler:2006qa} or some regions of the NMSSM \cite{Kim:2015dpa}.
\item {\em Parton showers versus simple decays.} If the mass of the LOSP (lightest ordinary superpartner) is much larger than the confinement scale in the hidden sector, we should think of its decays in terms of the unconfined theory, e.g.~${\tilde \chi}^0 \to g_h {\tilde g}_h$. The emitted hidden-sector gluinos and gluons can then radiate additional hidden gluons, leading to a high-multiplicity final state parton shower as illustrated in figure \ref{fig:partonshower}. Once the confinement scale of the hidden sector is reached, the many partons confine into a large number of bound states. On the other hand, if the mass of the LOSP is near the confinement scale of the hidden sector, the decay will involve just a few particles, e.g.~${\tilde \chi}^0 \to S {\tilde S}$ $(S' {\tilde S'})$, ${\tilde \chi}^0 \to Z{\tilde S}$, or ${\tilde \chi}^0 \to h{\tilde S}$, where $S$ ($S'$) and $\tilde S$ ($\tilde S'$) are bound states of hidden gluons (see section $3.3$).
\item {\em Light $SU(N)_h$ fundamentals or not.} If the particles $\Psi_i$ that transform as fundamentals of $SU(N)_h$ and have no Standard Model charge are light, the confining hidden theory can be QCD-like rather than a theory of pure glue. In particular, there may be light pion-like bound states. On the other hand, if all of the $\Psi$ particles have weak scale masses, there will only be bound states of hidden gluons and gluinos.
\item {\em Flavor-blind mediation or not.} The simplest ways of mediating SUSY breaking treat all the Standard Model flavors equally. This is appealing from the point of view of constraints on FCNCs, but due to direct search bounds on squarks (even in the scenario we consider) it pushes the theory into a somewhat fine-tuned regime. A variation could consider models that treat the third generation differently from the first two, as in Natural SUSY \cite{Cohen:1996vb,Dimopoulos:1995mi}.
\end{itemize}
In this paper we will consider one of the simplest possibilities: gaugino mediation gives a flavor-blind scheme that mediates to the SM and not the hidden sector, realizing Stealth SUSY. We work in a regime where the confinement scale is large enough compared to the LOSP mass that we can approximate the leading decays as involving a few particles rather than a parton shower. This case is illustrated in figure \ref{fig:nopartonshower}. Finally, because the mediation scheme we consider will tend to produce tachyonic scalars with $SU(N)_h$ charges if we try to arrange for light $SU(N)_h$ fundamental fermions, we consider the case of a pure-glue hidden valley. However, we emphasize that other choices in model-building can lead to different phenomenology. In particular, it would be appealing to construct a version of this idea realizing Natural SUSY, in which stops are light but first-generation squarks are much heavier. It would also be very interesting to consider the case where hidden-sector parton showers and jet physics play a role in the signals.

\begin{figure}[!h]
\begin{center}   
\scalebox{0.8}{
\begin{tikzpicture}[line width = 1.5 pt]
\node at (-1.0,-0.5) {\Large ${\tilde \chi}^0$};
\draw (-2.0,0.0)--(0.0,0.0);
\draw[dashed] (-4.0,0.0)--(-2.0,0.0);
\node at (-3.0,-0.5) {\Large ${\tilde q}$};
\draw (-2.0,0.0)--(-1.0,1.0); 
\node at (-0.9,1.5) {\Large $q$};
\node at (0.8,1.5) {\Large $g_h$};
\node at (0.8,-1.5) {\Large ${\tilde g}_h$};
\draw[gluon] (0,0)--(2.5,2.5);
\draw[gluon] (0.5,0.5)--(2.5,1.0);
\draw[gluon] (2.0,0.875)--(2.5,0.6);
\draw[gluon] (1,1)--(2.5,1.81);
\draw[gluon] (0,0)--(2.5,-2.5);
\draw (0,0)--(2.5,-2.5);
\draw[gluon] (0.5,-0.5)--(2.5,-0.1);
\draw[gluon] (1.5,-0.3)--(2.5,-0.8);
\draw[gluon] (1.5,-1.5)--(2.5,-1.4);
\fill[blue!50!white] (2.5,3.0) rectangle (3.0,-3.0);
\draw[dashed] (3.0,2.8)--(4.5,3.5); \node at (4.3,3.8) {\Large $S'$};
\draw[fermion] (4.5,3.5)--(5.5,4.0); \node at (5.8,4.0) {\Large $b$};
\draw[fermion] (5.5,3.0)--(4.5,3.5); \node at (5.8,3.0) {\Large $\bar b$};
\draw[dashed] (3.0,1.5)--(4.5,2.0); \node at (4.3,2.3) {\Large $S$};
\draw[fermion] (4.5,2.0)--(5.5,2.5); \node at (5.8,2.5) {\Large $b$};
\draw[fermion] (5.5,1.5)--(4.5,2.0); \node at (5.8,1.5) {\Large $\bar b$};
\draw[dashed] (3.0,-0.2)--(4.5,0.0); \node at (4.3,0.3) {\Large $S$};
\draw[fermion] (4.5,0.0)--(5.5,-0.2); \node at (5.8,-0.2) {\Large $b$};
\draw[fermion] (5.5,0.6)--(4.5,0.0); \node at (5.8,0.6) {\Large $\bar b$};
\draw[dashed] (3.0,-1.2)--(4.5,-1.2); \node at (4.3,-0.9) {\Large $S'$};
\draw[fermion] (4.5,-1.2)--(5.5,-1.8); \node at (5.9,-1.8) {\Large $\tau^-$};
\draw[fermion] (5.5,-0.8)--(4.5,-1.2);  \node at (5.9,-0.8) {\Large $\tau^+$};
\draw (3.0,-2.8)--(4.5,-3.5); \node at (3.7,-3.5) {\Large $\tilde S$};
\draw[dashed] (4.5,-3.5)--(5.5,-3.0); \node at (5.0,-2.9) {\Large $S$};
\draw[fermion] (6.5,-3.5)--(5.5,-3.0); \node at (6.8,-3.5) {\Large $\bar b$};
\draw[fermion] (5.5,-3.0)--(6.5,-2.5); \node at (6.8,-2.5) {\Large $b$};
\draw (4.5,-3.5)--(5.5,-4.0); \node at (6.0,-4.0) {\Large ${\tilde G}$};
\end{tikzpicture}
}
\end{center}
\caption{A squark decay process in the scenario with a parton shower preceding confinement. The vertical blue bar represents hidden-sector hadronization. As in Hidden Valley models, very large final-state multiplicities can arise if the confinement scale is sufficiently low compared to the mass of ${\tilde \chi}^0$. This is an interesting scenario that we will defer studying until a future paper.}
\label{fig:partonshower}
\end{figure}
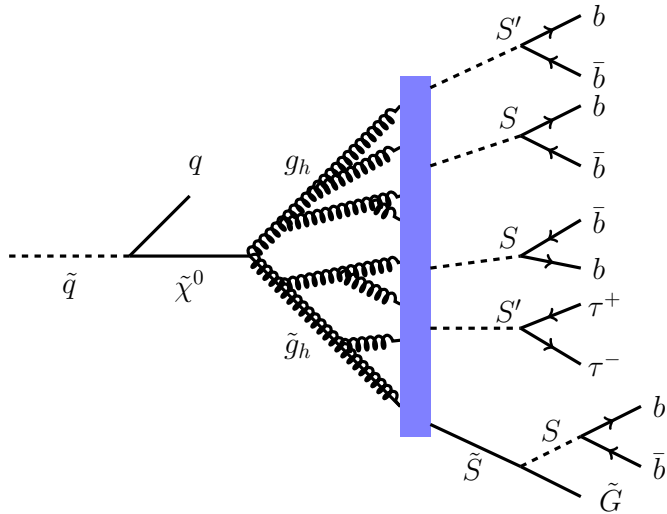

\subsection{The Hidden Valley spectroscopy}

To estimate the spectrum of confined states, we use results of lattice computations. Unfortunately, lattice results are not always presented in a manner that allows transparent comparison to perturbation theory.
By solving RGEs, we can obtain a perturbative estimate of the confinement scale $\Lambda^{\overline{MS}}$ in the $\overline{MS}$ scheme, as in equation (\ref{eq:runningalphah}).
On the other hand, the lattice can compute the spectrum of massive states of a theory in units of a nonperturbative quantity like the string tension $\sigma$
or the Sommer scale parameter $r_0$, defined by $F(r_0) r_0^2 = 1.65$ where $F(r)$ is the force at distance $r$ determined by the static potential \cite{Sommer:1993ce}.
Given masses quoted in such a nonperturbative scheme, we need to know how to match to perturbation theory, for instance by knowing the value of the dimensionless number $r_0 \Lambda^{\overline{MS}}$.
This requires a matching calculation that depends on the particular theory in question. For QCD-like theories with zero or two flavors,
the quantity $r_0 \Lambda^{\overline{MS}}$ has been computed and is $\approx 0.6$ in both cases~\cite{Gockeler:2005rv}. In the absence of such a matching calculation for a general theory,
we will quote masses based on the estimate $r_0 \Lambda^{\overline{MS}} \approx 0.6$ as well as a range $0.4 \leq r_0 \Lambda^{\overline{MS}} \leq 0.9$,
multiplying and dividing by 1.5 to capture possible variations in the matching for non-QCD-like theories.
Because our goal is to highlight the broad LHC signatures in a simplified model framework, rather than to give numerically precise details of the masses and couplings, this order-one uncertainty is acceptable.

We assume that the vectorlike fermions all have masses significantly larger than the confinement scale, $m', m'_0 \gg \Lambda$.
In this case we have a pure-glue Hidden Valley, either nonsupersymmetric or with approximate ${\cal N} = 1$ supersymmetry depending on the relative size of the gaugino mass $m_\lambda$
and the confinement scale $\Lambda$. The phenomenology of nonsupersymmetric pure-glue Hidden Valleys has been discussed in refs.~\cite{Juknevich:2009ji,Juknevich:2009gg},
building on lattice gauge theory results for the spectrum of pure Yang-Mills theory~\cite{Morningstar:1999rf,Chen:2005mg}. The lightest glueball is a $0^{++}$ state with mass $4.2 r_0^{-1}$,
translating to about $7 \Lambda^{\overline{MS}}$ (or, accounting for matching uncertainty, between about $4.7\Lambda^{\overline{MS}}$ and $11\Lambda^{\overline{MS}}$.
The next states have quantum numbers $2^{++}$ and mass $5.8 r_0^{-1}$; $0^{-+}$ and mass $6.3 r_0^{-1}$; and $1^{+-}$ with mass $7.3 r_0^{-1}$.
Due to the numerous closely spaced states with different quantum numbers, there are a large number of stable glueballs (in the absence of higher-dimension operators linking the Yang-Mills theory to other light particles that provide decay modes).

The ${\cal N}=1$ supersymmetric Yang-Mills spectrum has only recently begun to come under control on the lattice \cite{Bergner:2013nwa,Bergner:2013jia,Bergner:2014iea,Bergner:2015iva,Bergner:2015cqa},
with a reliable extrapolation to $m_\lambda = 0$ showing a mass-degenerate lightest supermultiplet as expected. 
This supermultiplet, which mostly overlaps the multiplet containing the ``gluinoball'' operator $\lambda \lambda$, has a mass of about $2.7 r_0^{-1}$,
while the heavier supermultiplet (mostly overlapping the ``glueball'' operator ${\rm Tr}\, G^2$) has a mass of about $3.3 r_0^{-1}$. These translate to $4.5\Lambda^{\overline{MS}}$ and $5.5\Lambda^{\overline{MS}}$
with the central estimate for $r_0 \Lambda^{\overline{MS}}$. Again, there is an order-one matching uncertainty attached to these numbers.

Notice that the glueball (and gluinoball) masses are, on a logarithmic scale, closer to $10 \Lambda^{\overline{MS}}$ than to $\Lambda^{\overline{MS}}$.
This is an important point: even if the RGE estimate is that the confinement scale is significantly below the scale of superpartner masses,
the actual masses of confined states may not be so light. If MSSM superpartners decay to hidden-sector particles that are confined, there may be relatively little room for a parton shower to produce high-multiplicity final states unless $\Lambda^{\overline{MS}}$ is quite low.


\subsection{The simplified model of Hidden Valley phenomenology}

We now present a simplified model of the Higgs and Hidden Valley fields for collider phenomenology which will be useful for later discussions.
Let us concentrate on two of the lightest supermultiplets, discussed above, containing the gluinoball and glueball operators.
The simplified model has two SM singlet chiral superfields denoted as $S$ and $S'$.
We can roughly identify $S$ as the gluinoball chiral superfield,
\begin{equation}
\begin{split}
S &\sim \tr \left( W^\alpha W_\alpha \right) / \Lambda^2,
\end{split}
\end{equation}
and $S'$ as the glueball chiral superfield whose lowest component is proportional to $\tr \left( F_{\mu\nu} F^{\mu\nu} \right)$.
The effective description of the gluinoball and glueball supermultiplets is still unclear although some attempts have been known
(see e.g.~\cite{Farrar:1997fn}).
Then, it is important to note that our simplified model does not mean the effective theory of the pure supersymmetric Yang-Mills theory after the confinement,
but this treatment is sufficient for our purpose.
The simplified model is useful when there is no high multiplicity of exotic new states in decays of ordinary sector particles.
The superpotential of our simplified model is given by
\begin{equation}
\begin{split}
W_{\rm simplified} &=  \mu H_u H_d + \lambda_S S H_u H_d + m_{SS'} S S' + \frac{1}{2} m_S S^2 + \frac{1}{2} m_{S'} S'^2
  \\
 &\quad + \frac{1}{3}\kappa S^3 + \left(\text{cubic terms with $S'$} \right) . \label{Wsimplified} \\
\end{split}
\end{equation}
The first term is the usual $\mu$ term and the second is the coupling between the Higgs and the Hidden Valley fields
generated after integrating out the vectorlike matter fields as we will see below.
The next three terms represent the supersymmetric masses of the gluinoball and glueball multiplets.
From the discussion in the previous subsection, we assume the sizes of the mass parameters as
\begin{equation}
\begin{split}
m_S \sim m_{S'} \sim 5 \Lambda. \label{susymass}
\end{split}
\end{equation}
The two multiplets $S$ and $S'$ mix significantly and thus we assume the mixed mass parameter $m_{SS'}$ is of the same order.
The coupling of the cubic term is estimated as $\kappa \sim 4 \pi $ by using Naive Dimensional Analysis (NDA)
\cite{Luty:1997fk}.

The hidden gluino mass induces SUSY-breaking mass splittings in the gluinoball and glueball supermultiplets.
The small gluino mass can be accommodated in terms of the $\theta^2$ component of the holomorphic coupling.
Since the confinement scale depends on the holomorphic coupling, $\Lambda$ also gets a $\theta^2$ component.
As in \eqref{susymass}, the supersymmetric mass parameters are determined by the confinement scale and they are expected to obtain
nonzero $F$ components.
Therefore, we here assume
\begin{equation}
\begin{split}
m_S \rightarrow m_S \left( 1+ \tilde{m}_{S} \theta^2 \right) , \qquad m_{S'} \rightarrow m_{S'} \left( 1+ \tilde{m}_{S'} \theta^2 \right) , \label{softmass}
\end{split}
\end{equation}
where $\tilde{m}_S \sim \tilde{m}_{S'}$ originally come from the gluino mass.
With these nonzero $F$ components, the squared masses of the singlet scalars are $m_S ( m_S \pm \tilde{m}_S)$ and $m_{S'} ( m_{S'} \pm \tilde{m}_{S'})$.
Here, we have ignored the second and third terms in the superpotential \eqref{Wsimplified}.
Note that one scalar is heavier and the other is lighter than the fermion in each supermultiplet.
When the hidden gluino mass is much larger than the confinement scale, this spurion argument is not appropriate.
However, the mass of the composite which contains the gluino as a constituent is almost determined by the gluino mass in this case.
Then, at least one scalar, the glueball, is lighter than the gluino-glue fermion.
Therefore, in the following discussions, we assume that one scalar is always lighter than the gluino-glue fermion.
For the Higgs fields, there are the usual soft terms such as the quadratic mass parameters $\tilde{m}_{H_u}^2$, $\tilde{m}^2_{H_d}$ and the $b_\mu$ term.
Although we can introduce $F$ components into the other terms in the superpotential,
we do not consider them just for simplicity.

The interaction strength between the Higgs sector and the singlet chiral superfields can be estimated by comparing amplitude calculations in terms of the gauge theory and the presented simplified model.
For the gauge theory side, the effective interactions between the Higgs sector and the new gauge fields are generated after integrating out the vectorlike fields.
We assume just for simplicity that $m \sim m'_0$ and all soft masses of the vectorlike fields are ${\tilde m}^2$.
Then, consider the following two cases: where the $\lambda_d$ coupling is sizable, $\lambda_u \simeq \lambda_d$, and the soft breaking terms of the vectorlike fields are small, $m^2 \gg \tilde{m}^2$;
where $\lambda_d$ is tiny, $\lambda_d \ll 1$, and the soft breaking terms are not small, $m^2 \gsim \tilde{m}^2$.

\subsubsection{The case with $\lambda_u \simeq \lambda_d$ and $m^2 \gg \tilde{m}^2$}

To know the effective interactions between the Higgs fields and the gauge fields,
we integrate out the vectorlike fields supersymmetrically.
The gauge coupling of the low-energy effective theory depends on the Higgs vevs
from which we can extract the effective interactions.
With the canonically normalized gauge kinetic term,
the coupling between the Higgs sector and the hidden gauge field is given by the dimension-six operator,
\begin{equation}
\begin{split}
\mathcal{L}_{\rm eff} &= - \frac{ig_h^2}{16\pi}\int d^2 \theta \, \tau(\mu) \, \tr \, W^\alpha W_\alpha + {\rm h.c.} \\
&= - \frac{g_h^2 \lambda_u \lambda_d}{32\pi^2 m m'_0} H_u H_d \, \tr \, F_{\mu\nu} F^{\mu\nu} + {\rm h.c.} + \cdots , \label{dimsix}
\end{split}
\end{equation}
where $\tau(\mu) = \frac{\theta_{\rm YM}}{2\pi} + \frac{4\pi i}{g_h^2}$ is the holomorphic coupling.
Note that the Higgs fields enter into the expression with the holomorphic combination $H_u H_d$.
We now consider the decay of the glueball/gluinoball scalar $0^{++}$ to a pair of SM particles
and compare the calculations in terms of the gauge theory and the simplified model.
Here, we use the standard definitions of the neutral components of the Higgs fields,
\begin{equation}
\begin{split}
H_u^0  =  \frac{1}{\sqrt{2}} \left( v \sin \beta + h \cos \alpha + \cdots + i a_u \right) , \\
H_d^0  =  \frac{1}{\sqrt{2}} \left( v \cos \beta - h \sin \alpha + \cdots  + i a_d \right) ,
\end{split}
\end{equation}
where $v \simeq 246 \, \rm GeV$ is the Higgs vev.
First, consider the gauge theory calculation.
Using the dimension-six operator \eqref{dimsix}, the decay amplitude of the $0^{++}$ state via $0^{++} \rightarrow h^\ast \rightarrow\zeta \zeta$ ($\zeta$ denotes a SM particle)
can be calculated as
\begin{equation}
\begin{split}
\frac{g_h^2 \lambda_u \lambda_d \, v}{32\pi^2 m m'_0} \cdot  \cos \left( \alpha + \beta \right) \cdot 
\langle \zeta \zeta | y_{\xi} \xi \bar{\xi} + \cdots | 0 \rangle \cdot \frac{1}{m_h^2 - m_{0^{++}}^2} \cdot 
F_{0^{++}} ,
\end{split}
\end{equation}
where $y_{\xi}$ is the Yukawa coupling of a SM fermion $\xi$, $F_{0^{++}}$ is the $0^{++}$ decay constant and $m_{0^{++}}$ is the mass of the lightest scalar $0^{++}$.
On the other hand, in the simplified model, the same decay amplitude can be estimated as
\begin{equation}
\begin{split}
\lambda_S m_{SS'} v \cdot \cos \left( \alpha + \beta \right) \cdot 
\langle \zeta \zeta | y_{\xi} \xi \bar{\xi} + \cdots | 0 \rangle \cdot \frac{1}{m_h^2 - m_{0^{++}}^2} , \label{amplitude}
\end{split}
\end{equation}
where we have used the scalar trilinear interaction,
$\mathcal{L}_{\rm simplified} \supset - \lambda_S m_{SS'} S' (H_u H_d )^\ast + {\rm h.c.}$,
in the simplified model Lagrangian.
We can replace the mass parameter $m_{SS'}$ to $m_S$ in this expression although the qualitative result is unchanged.
Then, comparing these two amplitudes, we obtain
\begin{equation}
\begin{split}
\lambda_S = \frac{g_h^2 \lambda_u \lambda_d}{32\pi^2} \frac{F_{0^{++}}}{m m'_0 m_{SS'}}.
\end{split}
\end{equation}
In the nonsupersymmetric case, the lattice result  \cite{Chen:2005mg} tells us that
$g_h^2 F_{0^{++}} = 3.06 \, m_{0^{++}}^3$.
When we assume this value of the decay constant, $\lambda_u = 1.5$, $\lambda_d = 1.0$, the masses of the vectorlike fields are $m = m'_0 = 300 \, \rm GeV$
and $m_{SS'} = m_{0^{++}} = 50 \, \rm GeV$, the $\lambda_S$ coupling is estimated as $\lambda_S \sim 0.4 \times 10^{-3}$.
It is important to note that this coupling is proportional to $\lambda_d$ in this case.
We concentrate on this case in the rest of the discussions of this section although we briefly look at another case just below.

\subsubsection{The case with $\lambda_d \ll 1$ and $m^2 \gsim \tilde{m}^2$}

When $\lambda_d$ is tiny and the soft breaking terms of the new vectorlike fields are not small, the effective interaction \eqref{dimsix} is negligible but
other interactions are generated after integrating out the vectorlike fields.
They include the effective interaction between the Higgs and the glueball scalar.
By using a SUSY-breaking spurion $\tilde{m}^2 \theta^4$, we can write the following operator at the leading order of the soft breaking parameter,
\begin{equation}
\begin{split}
& \int d^4 \theta \,  \frac{g_h^2 |\lambda_u|^2}{16 \pi^2} \frac{\tilde{m}^2 \theta^4}{|m|^2 |m'_0|^2}  H_u^\dagger H_u \left( {D}^2 \tr \, W^\alpha W_\alpha + {\rm h.c.} \right), \label{dimsixhard}
\end{split}
\end{equation}
where $D_\alpha$ is the superspace derivative.
The coefficient has been estimated by NDA. 
This operator includes the interaction $H_u^\dagger  H_u \tr \, F_{\mu\nu} F^{\mu\nu}$ which corresponds to the hard breaking term $S' H_u^\dagger H_u$ in the simplified model.
Note that the Higgs enters into the expression with the combination $H_u^\dagger H_u$ unlike the previous case.
We can estimate the size of the (dimensionful) coupling of $S' H_u^\dagger H_u$ by comparing the calculations of the decay of the glueball $0^{++}$ to a pair of SM particles
in terms of the gauge theory and the simplified model as before.
When we assume that $\lambda_u = 1.5$, the masses of the vectorlike fields are $m = m'_0 = 300 \, \rm GeV$, the soft breaking parameter is $\tilde{m}^2 = (300 \, {\rm GeV})^2$
and $m_{0^{++}} = 50 \, \rm GeV$, this coupling is estimated as $\sim 0.06 \, \rm GeV$.
Moreover, for the gluino-glue fermion, we find the effective interaction with the Higgs and higgsino.
At the leading order of SUSY breaking, we can write
\begin{equation}
\begin{split}
& \int d^4 \theta \, \frac{g_h^2 |\lambda_u|^2}{16 \pi^2}  \frac{\tilde{m}^2 \theta^4}{|m|^2 |m'_0|^2}  \left( H_u^\dagger D^\alpha H_u {D}_\alpha \tr \, W^\alpha W_\alpha + {\rm h.c.} \right).
\end{split}
\end{equation}
Note that this operator also does not include the down-type Higgs or higgsino and corresponds to the hard breaking term in the simplified model.
The interactions between the Higgs sector and the Hidden Valley fields have an important role in hiding supersymmetric particles at the LHC.

\subsection{The LOSP decay}\label{sec:LOSPdecay}

\begin{figure}[!h]
\begin{center}
\begin{tikzpicture}[line width = 1.5 pt]
\draw[fermion] (-1.0,0)--(0.25,0);
\node at (-0.5,-0.35) {${\tilde H}^0$};
\node at (1.0,-0.35) {${\tilde S}$};
\draw[dashed] (0.25,0)--(0.25,0.5);
\node at (0.25,0.5) {\Large $\times$};
\node at (0.25,0.95) {$\left<h\right>$};
\draw[fermion] (0.25,0)--(1.5,0);
\draw[dashed] (1.5,0)--(2.5,1.0);
\node at (2.65,1.0) {$S$};
\draw[fermion] (1.5,0)--(2.5,-1.0);
\node at (2.65,-1.0) {${\tilde S}$};
\begin{scope}[shift={(4.0,0)}]
\draw[fermion] (0,0)--(1.5,0);
\node at (0.75,-0.35) {${\tilde H}^0$};
\draw[dashed] (1.5,0)--(2.5,1.0);
\node at (2.65,1.0) {$h$};
\draw[fermion] (1.5,0)--(2.5,-1.0);
\node at (2.65,-1.0) {${\tilde S}$};
\end{scope}
\begin{scope}[shift={(8.0,0)}]
\draw[fermion] (0,0)--(1.5,0);
\node at (0.75,-0.35) {${\tilde H}^0$};
\draw[photon] (1.5,0)--(2.5,1.0);
\node at (2.7,1.0) {$Z$};
\draw[fermion] (1.5,0)--(2.25,-0.75);
\node at (1.75,-0.65) {${\tilde H}^0$};
\draw[dashed] (2.25,-0.75)--(2.6,-0.45);
\node at (2.6,-0.45) {\Large $+$};
\node at (3.1,-0.45) {$\left<h\right>$};
\draw[fermion] (2.25,-0.75)--(3.0,-1.5);
\node at (3.15,-1.5) {${\tilde S}$};
\end{scope}
\begin{scope}[shift={(12.0,0)}]
\draw[fermion] (0,0)--(1.5,0);
\node at (0.75,-0.35) {${\tilde H}^\pm$};
\draw[photon] (1.5,0)--(2.5,1.0);
\node at (2.85,1.0) {$W^\pm$};
\draw[fermion] (1.5,0)--(2.25,-0.75);
\node at (1.75,-0.65) {${\tilde H}^0$};
\draw[dashed] (2.25,-0.75)--(2.6,-0.45);
\node at (2.6,-0.45) {\Large $+$};
\node at (3.1,-0.45) {$\left<h\right>$};
\draw[fermion] (2.25,-0.75)--(3.0,-1.5);
\node at (3.15,-1.5) {${\tilde S}$};
\end{scope}
\end{tikzpicture}
\end{center}
\caption{Decays of the neutral higgsino to singlet plus singlino, higgs plus singlino, or $Z$ plus singlino, and of the charged higgsino to $W$ plus singlino.}
\label{fig:LOSPdecay}
\end{figure}
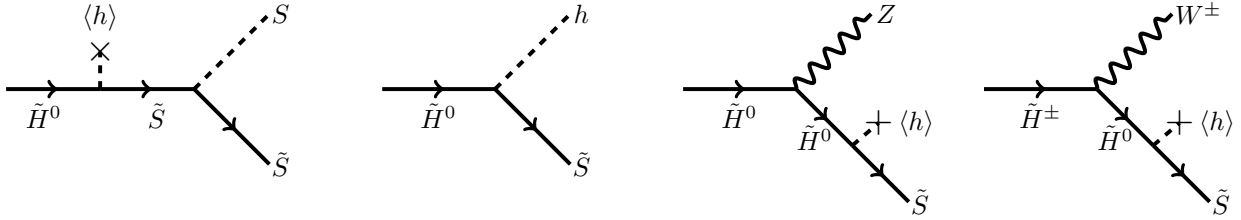

In our framework, we assume low-scale SUSY breaking at $10-100 \, \rm TeV$, and the lightest supersymmetric particle (LSP) is the gravitino.
The lightest ordinary supersymmetric particle (LOSP) is assumed to be the higgsino-like neutralino $\tilde{\chi}_1^0$ and then
the LOSP decays to exotic new states through the interactions between the Higgs and Hidden Valley sectors, as depicted in figure \ref{fig:LOSPdecay}. Details of such decays arising through a superpotential interaction $\lambda_S S H_u H_d$ are computed in ref.~\cite{Stealth3}. However, in our case there is an important difference: the composite states are strongly coupled to each other through operators like $\kappa S^3$. This means that the decay ${\tilde \chi}^0 \to S {\tilde S}$ will be the dominant decay process: it proceeds through the higgsino--singlino mixing and then a coupling of order $4\pi$. This distinguishes the decays in our scenario from those in other Stealth SUSY $SH_u H_d$ models considered in the past that had small values of $\kappa$ \cite{Fan:2011yu,Fan:2012jf,Stealth3}, for which this decay is usually subdominant or at most an order-one fraction of the decays. However, the case where ${\tilde \chi}^0 \to S{\tilde S}$ dominates has been studied in ref.~\cite{Evans:2013jna}.

Higgsinos come in a nearly-degenerate multiplet: ${\tilde \chi}^0_1$, ${\tilde \chi}^\pm_1$, and ${\tilde \chi}^0_2$ have small mass splittings $\sim m_Z^2/M_2$, which may be $5-10\, \rm GeV$ for $M_2 \sim {\rm TeV}$ and $\mu \sim 100\, {\rm GeV}$. As a result, even the heavier states in the multiplet may decay directly to the singlino rather than to the LOSP: ${\tilde \chi}^\pm \to W^\pm {\tilde S}$ pays the price of a small coupling $\lambda_S$ but ${\tilde \chi}^\pm \to W^{*\pm} {\tilde \chi}^0_1$ is a three-body decay that is highly phase-space suppressed.

The decay width of $\tilde\chi_1^0 \to h \tilde S$ and $\tilde\chi_1^0 \to S \tilde S$ is estimated as
\begin{equation}
\begin{split}
\Gamma_{\tilde{\chi}_1^0 \rightarrow h \tilde{S}} \, &\sim \,
\frac{\lambda_S^2 }{16\pi} \, m_{\tilde{\chi}_1^0} \left( 1 - \frac{m_h^2}{m_{\tilde{\chi}_1^0}^2} \right)  ,\\
\Gamma_{\tilde{\chi}_1^0 \rightarrow S \tilde{S}} \, &\sim \,
\frac{\lambda_S^2 }{16\pi} \, \frac{\kappa^2 v^2}{m_{\tilde{\chi}_1^0}},
\end{split}
\end{equation}
where $m_{\tilde{\chi}_1^0}$ is the lightest neutralino mass and the hidden gluino-glue mass is ignored.
In our setup, $\kappa $ is assumed to be $\sim 4\pi$ and $m_{\tilde\chi_1^0}$ is ${\cal O}(100)$ GeV.
Thus, the dominant decay of the LOSP is given by $\tilde\chi_1^0 \to S \tilde S, S \tilde S', S' \tilde S, S' \tilde S'$.
With a typical size of the $\lambda_S$ coupling, the decay is prompt.
When the confinement scale is large, the Higgs or the gluino-glue fermion becomes offshell in this decay process.
In this case, the width gets a phase space suppression.
The produced gluino-glue fermion decays to the gravitino and the lighter glueball/gluinoball scalar which decays back to SM particles as we will see next.

\subsection{The hidden glueball/gluinoball decays}\label{sec:glueball_decay}

Let us now consider decays of the hidden glueball and gluinoball scalars to a pair of SM particles.
They decay through the interactions between the Higgs and the Hidden Valley fields.
For the $0^{++}$ states, we have already estimated the amplitude in terms of the simplified model as in \eqref{amplitude}.
Then, the width of the $0^{++}$ decay to a pair of SM particles is denoted as
\begin{equation}
\begin{split}
\Gamma_{0^{++} \rightarrow \zeta \zeta} \sim 
\left( \frac{\lambda_S m_{SS'}v}{m_h^2 - m_{0^{++}}^2 }\right)^2
\Gamma_{h \rightarrow \zeta \zeta}^{\rm SM} (m_{0^{++}}^2)  ,
\end{split}
\end{equation}
where $\Gamma_{h \rightarrow \zeta \zeta}^{\rm SM} (m_{0^{++}}^2)$ is the width of the SM Higgs boson decay $h \rightarrow \zeta \zeta$ in the case that the Higgs mass is given by the $0^{++}$ mass.
The interesting point of this expression is that the branching fractions of the $0^{++}$ decays are the same with
those of the Higgs boson decays.
The total width of the Higgs boson with the mass of the $0^{++}$ state is $\sim 1.5 \, \rm MeV$ when we take $m_{SS'} = m_{0^{++}} = 50 \, \rm GeV$
\cite{Djouadi:2005gi}.
Then, with $\lambda_S = 10^{-3}$, the decay length of the $0^{++} $states is estimated as $c \tau_{0^{++}} \sim 0.1 \, \mu$m, which is not so displaced to be observed at the LHC.
The hidden glueball and gluinoball scalars can decay into a pair of SM gauge bosons through loops of the vectorlike fermions.
However, since the decay width is proportional to a high power of the confinement scale and suppressed by the vectorlike masses,
this mode is subleading in the present scenario where the confinement scale is much smaller than the vectorlike masses.

While the pseudoscalars $0^{-+}$ cannot decay through the dimension-six operator in the nonsupersymmetric theory
\cite{Juknevich:2009gg},
they are possible in the present supersymmetric theory through the CP odd Higgs boson as $0^{-+} \rightarrow A^\ast \rightarrow Z h, \tau^+ \tau^-, \mu^+ \mu^-, b \bar{b}$.
The decay width is estimated as
\begin{equation}
\begin{split}
\Gamma_{0^{-+} \rightarrow \zeta \zeta} \sim \left( \frac{\lambda_S m_{SS'} \, v}{ m_A^2 - m_{0^{-+}}^2 } \right)^2
\Gamma_{A \rightarrow \zeta \zeta}^{\rm MSSM} (m_{0^{-+}}^2) ,
\end{split}
\end{equation}
by using the interaction terms,
$\mathcal{L}_{\rm simplified} \supset - \lambda_S m_{SS'} S' (H_u H_d )^\ast + {\rm h.c.} =  - \lambda_S m_{SS'} v \, \eta_{S'} A + \cdots $,
where we have defined $S' = \xi_{S'} + i \eta_{S'}$ (The imaginary component $\eta_{S'}$ denotes $0^{-+}$)
and $A = a_u \cos \beta + a_d \sin \beta$
is the physical CP odd component of the two Higgs doublet model.
The branching fractions of the $0^{-+}$ decays are the same with
those of the CP odd Higgs boson decays.
The total width of the CP odd Higgs boson with the mass of the $0^{-+}$ state is given by $\sim 2 \, \rm GeV$ for $m_{SS'} = m_{0^{-+}} = 100 \, \rm GeV$ and $\tan \beta = 30$
\cite{Djouadi:2005gj}.
Then, the decay length is estimated as $c \tau_{0^{-+}} \sim 1 \, \rm nm$
where we have taken $m_{A} = 300 \, \rm GeV$.
The decay into a pair of SM gauge bosons through loops of the vectorlike fermions is suppressed as discussed above.

\subsection{The decays of gluino-glue fermions}\label{sec:glueballino_decay}

The lightest hidden gluino-glue fermion decays to the gravitino LSP and the glueball or gluinoball scalar which decays to a pair of SM particles as discussed above.
When the hidden gluino mass is small, the mass splitting between the scalar and the fermion in the glueball or gluinoball supermultiplet is also tiny.
In this case, the missing energy is reduced, which contributes to hiding supersymmetry at the LHC as proposed in Stealth Supersymmetry \cite{Fan:2011yu,Fan:2012jf}.
The decay width is given by
\begin{equation}
\begin{split}
\Gamma_{\tilde{S} \rightarrow S \tilde{G}} =
\frac{m_{\tilde{S}}^5}{16\pi F^2} \left( 1 - \frac{m_S^2}{m_{\tilde{S}}^2} \right)^4 \simeq \frac{m_{\tilde{S}} (\delta m)^4}{\pi F^2}
\simeq \frac{1}{1.4~{\rm mm}} \left( \frac{m_{\tilde S}}{50~{\rm GeV}} \right) \left( \frac{\delta m}{20~{\rm GeV}} \right)^4 \left( \frac{m_{3/2}}{1~{\rm eV}} \right)^{-2},
\end{split}
\end{equation}
where $\sqrt{F}$ is the SUSY-breaking scale
and $\delta m$ is the mass splitting between the scalar and the fermion,
both of which suppress the decay width.
The gravitino mass is given by $m_{3/2} = F/\sqrt{3}M_{\rm Pl}$.
If there is some mass hierarchy between the two gluino-glue fermions, the heavier fermion possibly decays to the lighter one and a pair of SM particles through the offshell Higgs.
The dominant mode is $\tilde{S}_2 \rightarrow \tilde{S}_1 h^\ast \rightarrow \tilde{S}_1  \zeta \zeta$ where $\tilde{S}_1$ and $\tilde{S}_2$ are the lighter and heavier fermions respectively.
The decay width can be estimated as
\begin{equation}
\begin{split}
\Gamma_{\tilde{S}_2 \rightarrow \tilde{S}_1 \zeta\zeta } \sim \frac{ (4\pi)^2\lambda_S^2 v^2 \delta m_{\tilde{S}}^2}{128 \pi^3(m_{\tilde{S}_2}^2 - m_h^2)^2} \, \Gamma_{h \rightarrow \zeta \zeta}^{\rm SM} (\delta m_{\tilde{S}}^2),
\end{split}
\end{equation}
where $\delta m_{\tilde{S}} = m_{\tilde{S}_2} - m_{\tilde{S}_1}$ and $m_{\tilde{S}_1}$, $m_{\tilde{S}_2}$ are the lighter and heavier gluino-glue fermion masses.
We have used the cubic term in the superpotential of the simplified model \eqref{Wsimplified}.
When we take $m_{\tilde{S}_1} = 50 \, \rm GeV$, $m_{\tilde{S}_2} = 60 \, \rm GeV$ and $\lambda_S = 10^{-3}$,
the widths of the heavier gluino-glue decay to the gravitino and the lighter gluino-glue are of similar order.
Therefore, the heavier gluino-glue fermion decays to the lighter one with a pair of SM particles as well as the gravitino.
In the rest of discussions, we assume that the mass splitting between the two gluino-glue fermions is small and
do not consider this decay mode.

\subsection{The effect on Higgs decays}

Decays of the SM-like Higgs boson in our framework may deviate from those of the SM.
First, we consider the Higgs decay to two photons, $h \rightarrow \gamma \gamma$.
Particles with SM electroweak quantum numbers coupling to the Higgs boson potentially induce
measurable changes to the Higgs branching ratio to two photons through their loop effects.
However, there are no mass terms of the electrically charged vectorlike fields which depend on the Higgs vev.
Therefore, there are no important contributions to  $h \rightarrow \gamma \gamma$ from new exotic particles in the present setup.

Next, we look at the Higgs decay to a pair of (offshell or onshell) Hidden Valley particles.
The branching fraction of $h \rightarrow 0^{++} 0^{++}$ is constrained by the global fit of the signal strength of $h\rightarrow \gamma\gamma, W^+W^-, ZZ, b\bar b$ and $\tau\bar\tau$ indirectly.
Since we assume that the width of the Higgs decay to SM particles is the same as that of the SM, ${\rm Br}(h\rightarrow 0^{++} 0^{++}) < 0.19$ should be satisfied at 95\% C.L.~\cite{Belanger:2013xza}.
The Higgs decay to two glueball or gluinoball scalars is given by the interaction terms,
$\mathcal{L}_{\rm simplified} \supset - \frac{1}{2} \lambda_S \kappa' S'^2 (H_u H_d )^\ast + {\rm h.c.}$,
derived from the cubic term, $(\kappa'/2) S S'^2$, in the superpotential of the simplified model.
The branching fraction of $h \rightarrow 0^{++} 0^{++}$ is denoted as
\begin{equation}
\begin{split}
{\rm Br} (h \rightarrow 0^{++} 0^{++})  &= 
\frac{\Gamma_{h \rightarrow 0^{++} 0^{++} }}{\Gamma_h^{\rm SM}
+ \Gamma_{h \rightarrow 0^{++} 0^{++} }},
\end{split}
\end{equation}
where $\Gamma_h^{\rm SM} = 4.41~{\rm MeV}$ for $m_h = 125~{\rm GeV}$ which is calculated by \texttt{HDECAY} \cite{Djouadi:1997yw}.
In this expression, the decay rate of the Higgs to two onshell glueball or gluinoball scalars is given by
\begin{equation}
\begin{split}
\Gamma_{h \rightarrow 0^{++} 0^{++}} \sim
\frac{|\lambda_S \kappa' v|^2}{16 \pi m_h} \sqrt{1 - \left(  \frac{2 m_{0^{++}}}{m_h}\right)^2} .
\end{split}
\end{equation}
Then, we obtain
${\rm Br} (h \rightarrow 0^{++}0^{++}) \sim 0.17$
when we take $\kappa' = 4 \pi$, $\lambda_S = 10^{-3}$ and $m_{0^{++}} = 50 \, \rm GeV$.
Although direct probes of $h\rightarrow 0^{++} 0^{++}$ are also possible, the present bound is not so strong.
The glueball or gluinoball scalar $0^{++}$ mainly decays into a pair of bottom quarks via mixing of the SM Higgs boson.
Therefore, the dominant exotic mode is $h\rightarrow 0^{++} 0^{++} \to 4b$.
However, there are large QCD backgrounds and no limits exist at present.
The decay to $2b 2\tau$ is possible but there seem to be no experimental searches for this mode.
The LHC multilepton searches weakly constrain the branching ratio of the Higgs decay to $4 \tau$ as ${\rm Br} ({h \rightarrow 4 \tau}) \lesssim 20 \%$ \cite{Curtin:2013fra}.
Therefore, we concentrate on the decay, $h \rightarrow 0^{++} 0^{++} \rightarrow b\bar{b} \mu^+\mu^-$ where $0^{++}$ is onshell.
The decay to offshell glueball or gluinoball scalars is too suppressed to be observed.
Then, the branching ratio of the decay mode  $h \rightarrow b\bar{b} \mu^+\mu^-$ is given by
\begin{equation}
\begin{split}
{\rm Br} ({h \rightarrow b\bar{b} \mu^+\mu^-})  &= 
2 \cdot
{\rm Br} ({h \rightarrow 0^{++} 0^{++}})
\cdot {\rm Br} ({0^{++} \rightarrow b\bar{b}})
\cdot {\rm Br} ({0^{++} \rightarrow \mu^+\mu^-}) .
\end{split}
\end{equation}
The branching fractions of the $0^{++}$ decays are given by
those of the Higgs boson decays by taking the Higgs boson mass as $m_{0^{++}}$.
Then, we obtain
${\rm Br} ({h \rightarrow b\bar{b} \mu^+\mu^-}) \sim 7 \times 10^{-5}$
when we take $\kappa' = 4 \pi$, $\lambda_S = 10^{-3}$ and $m_{0^{++}} = 50 \, \rm GeV$.
Here, ${\rm Br}(0^{++} \to b\bar b) = 0.87$ and ${\rm Br}(0^{++} \to \mu^+ \mu^-) = 2.4 \times 10^{-4}$ are calculated by \texttt{HDECAY} \cite{Djouadi:1997yw}.
Due to smallness of the branching ratio ${\rm Br} ({0^{++} \rightarrow \mu^+\mu^-})$, this is lower than the projected upper bound that could be achieved with Run 1 LHC data, ${\rm Br} ({h \rightarrow b\bar{b} \mu^+\mu^-}) \lesssim 10^{-4}$ \cite{Curtin:2013fra}.
The sensitivity to this channel is expected to reach few $\times~10^{-5}$ at 14 TeV LHC with ${\cal L} = 3000~{\rm fb}^{-1}$ \cite{Curtin:2014pda}.

Although these exotic decays of the Higgs boson are not yet very constrained by current data, they will be a very important probe of the scenario during  the LHC's Run 2 that is complementary to direct searches for superpartners.

\section{RGEs and benchmarks}
\label{sec:rge}

In this section we give detailed numerical results for some benchmark models by using the SARAH codes
\cite{Staub:2008uz,Staub:2013tta}.
As the initial condition of SUSY breaking, we consider the situation where the Hidden Valley sector is supersymmetric at the mediation scale.
This can be realized by low-scale gauge mediation \cite{Giudice:1998bp} with only SM charged messengers or low-scale gaugino mediation
\cite{Kaplan:1999ac,Chacko:1999mi}.
We here focus on gaugino mediation
from a relatively low scale, that is, we assume nonzero masses for the MSSM gauginos and vanishing scalar masses.
The initial gaugino masses have to be large enough for the scalar superpartners of the SM fermions to avoid the current LHC bound.
The Hidden Valley sector is supersymmetric and the hidden gluino mass is zero at the mediation scale.
The soft SUSY-breaking masses of the hidden gluino and the new vectorlike scalar fields are generated by the renormalization group effects.
At one-loop level, the terms including the MSSM gaugino masses or the new Yukawa couplings give dominant contributions to the vectorlike scalar masses,
\begin{align}
\frac{d}{d\log\mu} {\tilde m}_f^2 \simeq 
\frac{d}{d\log\mu} {\tilde {\bar m}}_f^2 &\simeq -\frac{32}{3} \frac{g_3^2}{16\pi^2}|M_3|^2 -\frac{32}{15} \frac{g_1^2}{16\pi^2}|M_1|^2, \\
\frac{d}{d\log\mu} {\tilde m}_u^2 &\simeq -\frac{6g_2^2}{16\pi^2}|M_2|^2 -\frac{6}{5} \frac{g_1^2}{16\pi^2}|M_1|^2 + \frac{2\lambda_d^2}{16\pi^2}( m_{H_d}^2 + {\tilde m}_u^2 + {\tilde {\bar m}}_0^2 ), \\
\frac{d}{d\log\mu} {\tilde m}_d^2 &\simeq -\frac{6g_2^2}{16\pi^2}|M_2|^2 -\frac{6}{5} \frac{g_1^2}{16\pi^2}|M_1|^2 + \frac{2\lambda_u^2}{16\pi^2}( m_{H_u}^2 + {\tilde m}_d^2 + {\tilde m}_0^2 ), \\
\frac{d}{d\log\mu} {\tilde {\bar m}}_0^2 &\simeq \frac{4\lambda_d^2}{16\pi^2}( m_{H_d}^2 + {\tilde m}_u^2 + {\tilde {\bar m}}_0^2 ), \\
\frac{d}{d\log\mu} {\tilde { m}}_0^2 &\simeq \frac{4\lambda_u^2}{16\pi^2}( m_{H_u}^2 + {\tilde m}_d^2 + {\tilde m}_0^2 ),
\end{align}
while there is no such contribution to $\tilde m^2$ and $\tilde{\bar m}^2$.
For these masses, the two-loop effect is important,
\begin{align}
\frac{d}{d\log\mu} {\tilde m}^2 \simeq 
\frac{d}{d\log\mu} {\tilde {\bar m}}^2 &\simeq \frac{c g_h^4}{(16\pi^2)^2} \left( 3({\tilde m}_f^2 + {\tilde{\bar m}}_f^2) + 2({\tilde m}_u^2 + {\tilde m}_d^2) + {\tilde m}_0^2 + {\tilde {\bar m}}_0^2 + (F-1)({\tilde m}^2 + {\tilde{\bar m}}^2 ) \right),
\label{singletRGE}
\end{align}
where $g_h$ is the hidden gauge coupling and $c$ is some numerical constant.
This two-loop contribution is included in all of the soft masses of the vectorlike scalar fields and
is not negligible because $g_h$ is large in our setup.
In fact, if $\lambda_d \ll 1$, the above term becomes dominant in the RG equation of ${\tilde{\bar m}}_0^2$.
On the other hand, the one-loop beta function of the hidden gluino mass is proportional to itself which is zero at the mediation scale.
However, the two-loop beta function includes the following term which is proportional to the MSSM gaugino masses:
\begin{align}
\frac{d}{d\log\mu} M_\lambda \simeq \frac{g_h^2}{(16\pi^2)^2}( {\tilde c}_1 g_1^2 M_1 + {\tilde c}_2 g_2^2 M_2 + {\tilde c}_3 g_3^2 M_3),
\end{align}
where $\tilde c_i$'s are some numerical constants.
Thus, the hidden gaugino mass is generated from the two-loop level but suppressed compared to the MSSM gaugino masses.\footnote{
There are also threshold corrections to the hidden gaugino mass from the vectorlike particles,
which are not larger than the running corrections.}

\renewcommand{\arraystretch}{1.3}
\begin{table}[t]
\begin{center}
\begin{tabular}{|c|c|c|c|c|}
 \hline
 & (A) & (B) & (C) & (D)
 \\
 \hline
$M_{1}$ [GeV]               & 2444 & 2216 & 2488 & 2259 \\ \hline
$M_{2}$ [GeV]               & 2483 & 2259 & 2527 & 2302 \\ \hline
$M_{3}$ [GeV]               & 2593 & 2380 & 2637 & 2422 \\ \hline
$-\sqrt{|m_{H_u}^2|}$ [GeV] & $-121$ & $-290$ & $-117$ & $-291$ \\ \hline
$m_{H_d}$ [GeV]             & 672 & 671 & 684 & 684 \\ \hline
$m_{q_1}$ [GeV]             & 1583 & 1585 & 1611 & 1615 \\ \hline
$m_{u_1}$ [GeV]             & 1484 & 1486 & 1511 & 1515 \\ \hline
$m_{d_1}$ [GeV]             & 1467 & 1469 & 1493 & 1497 \\ \hline
$m_{q_3}$ [GeV]             & 1552 & 1548 & 1579 & 1577 \\ \hline
$m_{u_3}$ [GeV]             & 1412 & 1401 & 1437 & 1427 \\ \hline
$m_{d_3}$ [GeV]             & 1467 & 1469 & 1493 & 1497 \\ \hline
$m_{l}$ [GeV]               & 682 & 682 & 694 & 695 \\ \hline
$m_{e}$ [GeV]               & 421 & 421 & 429 & 429 \\ \hline
\end{tabular}
\hspace{0.5cm}
\begin{tabular}{|c|c|c|c|c|}
 \hline
 & (A) & (B) & (C) & (D)
 \\
 \hline
$M_{\lambda}$ [GeV]                     & 45 & 49 & 46 & 50 \\ \hline
$\tilde{m}_{u}$ [GeV]                   & 625 & 621 & 630 & 626 \\ \hline
$\tilde{m}_{d}$ [GeV]                   & 615 & 612 & 621 & 617 \\ \hline
$\tilde{m}_{f}$ [GeV]                   & 1394 & 1395 & 1411 & 1413 \\ \hline
$\tilde{\bar{m}}_{f}$ [GeV]             & 1394 & 1395 & 1411 & 1413 \\ \hline
$-\sqrt{|\tilde{m}_{0}^2|}$ [GeV]       & $-302$ & $-310$ & $-311$ & $-321$ \\ \hline
$-\sqrt{|\tilde{\bar{m}}_{0}^2|}$ [GeV] & $-208$ & $-222$ & $-220$ & $-236$ \\ \hline
$-\sqrt{|\tilde{m}^2|}$ [GeV]           & $-180$ & $-194$ & $-193$ & $-210$ \\ \hline
$-\sqrt{|\tilde{\bar{m}}^2|}$ [GeV]     & $-180$ & $-194$ & $-103$ & $-210$ \\ \hline
$m_h$ [GeV]                             & 124.9 & 124.9 & 125.2 & 125.2\\ \hline
$\Lambda^{\overline{MS}} $ [GeV]        & 10 & 10 & 10 & 10 \\ \hline
$\lambda_u $                            & 1.42 & 1.42 & 1.42 & 1.42 \\ \hline
$M_{\rm m}$ [TeV]                       & 50 & 100 & 50 & 100 \\ \hline
\end{tabular}

\end{center}
\caption{Four benchmark model points (A), (B), (C), (D).
For (A) and (B), we take the number of the singlets as $F=2$ while $F=3$ for (C) and (D).
The mediation scale $M_m$ is taken to be $50 \, \rm TeV$ and $100 \, \rm TeV$ for (A), (C) and (B), (D) respectively.
The MSSM gaugino masses at the mediation scale are $M_1 = M_2 = M_3 = 2750,~2550,~2800,~2600 \rm GeV$ for (A), (B), (C), (D), respectively.
The $\lambda_d$ coupling is 0.5.
The confinement scale, which determines the hidden gauge coupling at the low scale, is taken to be $10 \, \rm GeV$ for all the points.
The table shows the numerical results of the Bino, Wino and gluino masses $M_1$, $M_2$, $M_3$, the (tachyonic) up-type Higgs soft mass $-\sqrt{|m_{H_u}^2|}$, the down-type Higgs soft mass $m_{H_d}$,
the 1st generation squark masses $m_{q_{1}}$, $m_{\bar{u}_{1}}$, $m_{\bar{d}_{1}}$ (the 2nd generation squark masses are almost the same),
the 3rd generation squark masses $m_{q_{3}}$, $m_{\bar{u}_{3}}$, $m_{\bar{d}_{3}}$, the slepton masses $m_{l}$, $m_{\bar{e}}$.
The table also shows the hidden gaugino mass $M_{\lambda}$,
the scalar masses of the new vectorlike pair of doublets $\tilde{m}_u$, $\tilde{m}_d$ and triplets $\tilde{m}_f$, $\tilde{\bar{m}}_f$,
the (tachyonic) scalar masses of the vectorlike pair of SM singlets which couples to the Higgs $\sqrt{|\tilde{m}_{0}^2|}$, $\sqrt{|\tilde{\bar{m}}_{0}^2|}$,
and the (tachyonic) scalar masses of the vectorlike pair of SM singlets without Higgs couplings $\sqrt{|\tilde{m}^2|}$, $\sqrt{|\tilde{\bar{m}}^2|}$.
For all three cases, the correct Higgs mass is obtained.
To avoid spontaneous breaking of the hidden gauge group, we take the supersymmetric mass parameters of the singlets as $m'_0 = m' = 700 \, \rm GeV$.
The other supersymmetric masses of the vectorlike fields are also $m = M = 700 \, \rm GeV$.}
\label{tab:spectra}
\end{table}
\renewcommand{\arraystretch}{1}

In Table~\ref{tab:spectra}, we show four benchmark model points (A), (B), (C), (D) to study in more detail at colliders below.
For (A) and (B), we take the number of the singlets as $F=2$ while $F=3$ for (C) and (D).
The mediation scale $M_m$ is taken to be $50 \, \rm TeV$ and $100 \, \rm TeV$ for (A), (C) and (B), (D) respectively.
The MSSM gaugino masses at the mediation scale are $M_1 = M_2 = M_3 = 2100 \, \rm GeV$.
The $\lambda_d$ coupling is 0.5.
The confinement scale, which determines the hidden gauge coupling at the low scale, is taken to be $10 \, \rm GeV$ for all the points.
The table shows the numerical results of the Bino, Wino and gluino masses $M_1$, $M_2$, $M_3$, the (tachyonic) up-type Higgs soft mass $-\sqrt{|m_{H_u}^2|}$, the down-type Higgs soft mass $m_{H_d}$,
the 1st generation squark masses $m_{q_{1}}$, $m_{\bar{u}_{1}}$, $m_{\bar{d}_{1}}$ (the 2nd generation squark masses are almost the same),
the 3rd generation squark masses $m_{q_{3}}$, $m_{\bar{u}_{3}}$, $m_{\bar{d}_{3}}$, the slepton masses $m_{l}$, $m_{\bar{e}}$.
The table also shows the hidden gaugino mass $M_{\lambda}$,
the scalar masses of the new vectorlike pair of doublets $\tilde{m}_u$, $\tilde{m}_d$ and triplets $\tilde{m}_f$, $\tilde{\bar{m}}_f$,
the (tachyonic) scalar masses of the vectorlike pair of SM singlets which couples to the Higgs $\sqrt{|\tilde{m}_{0}^2|}$, $\sqrt{|\tilde{\bar{m}}_{0}^2|}$,
and the (tachyonic) scalar masses of the vectorlike pair of SM singlets without Higgs couplings $\sqrt{|\tilde{m}^2|}$, $\sqrt{|\tilde{\bar{m}}^2|}$.
Due to the supersymmetric initial condition of the hidden gauge sector,
the hidden gluino mass and the soft scalar masses of the new vectorlike fields coupling to the Higgs are relatively small, but
for all four cases, the correct Higgs mass is obtained because the new Yukawa coupling is sizable as we discussed before.
However, some of the scalar soft masses of the vectorlike fields are tachyonic.
This can be seen from \eqref{singletRGE} where a sum rule among the scalar masses of the vectorlike fields is satisfied at low energies.
The SM charged vectorlike scalar masses always get positive contributions from the MSSM gaugino masses at one-loop level,
Then, the singlet scalar masses are driven to tachyonic.
To avoid spontaneous breaking of the hidden gauge group, we take the supersymmetric mass parameters of the singlets as $m'_0 = m' = 500 \, \rm GeV$.
The other supersymmetric masses of the vectorlike fields are also $m = M = 500 \, \rm GeV$.
For all the points in Table~\ref{tab:spectra}, the new Yukawa coupling $\lambda_u$ and the gauge coupling $g_h$ do not hit a Landau pole until at least $10^6 \, \rm GeV$.
The contributions to the $S$ and $T$ parameters from the new vectorlike fields are given by
\cite{Martin:2009bg}
\begin{equation}
\begin{split}
&\Delta S \approx 0.02 \times \left( \frac{N}{3} \right) \left( \frac{\lambda_u}{1.2} \right)^2
\left( \frac{500 \, \rm GeV}{m} \right)^2 \, , \\[1ex]
&\Delta T \approx 0.13 \times \left( \frac{N}{3} \right) \left( \frac{\lambda_u}{1.2} \right)^4 \left( \frac{500 \, \rm GeV}{m} \right)^2 \, ,
\end{split}
\end{equation}
where we have assumed $m = m'_0$ and $\sin \beta \approx 1$.
They are within the experimental bound at 95\% CL.
If we have another contribution to the $S$ parameter, constraints on $\lambda_u$ and $m$ are more relaxed.

Let us comment on fine tuning for the electroweak breaking in the present scenario.
We have assumed a low mediation scale to realize the mass hierarchy between the scalar masses and the MSSM gaugino masses.
As the mediation scale is larger, the hierarchy vanishes.
While the tachyonic up-type Higgs soft mass is driven by the stop mass,
the large Bino and Wino masses give a positive one-loop contribution to the up-type Higgs mass whose absolute value at the electroweak scale is reduced.

\section{Collider phenomenology}
\label{sec:collider}
\begin{figure}
\centering
\includegraphics[width=0.7\hsize]{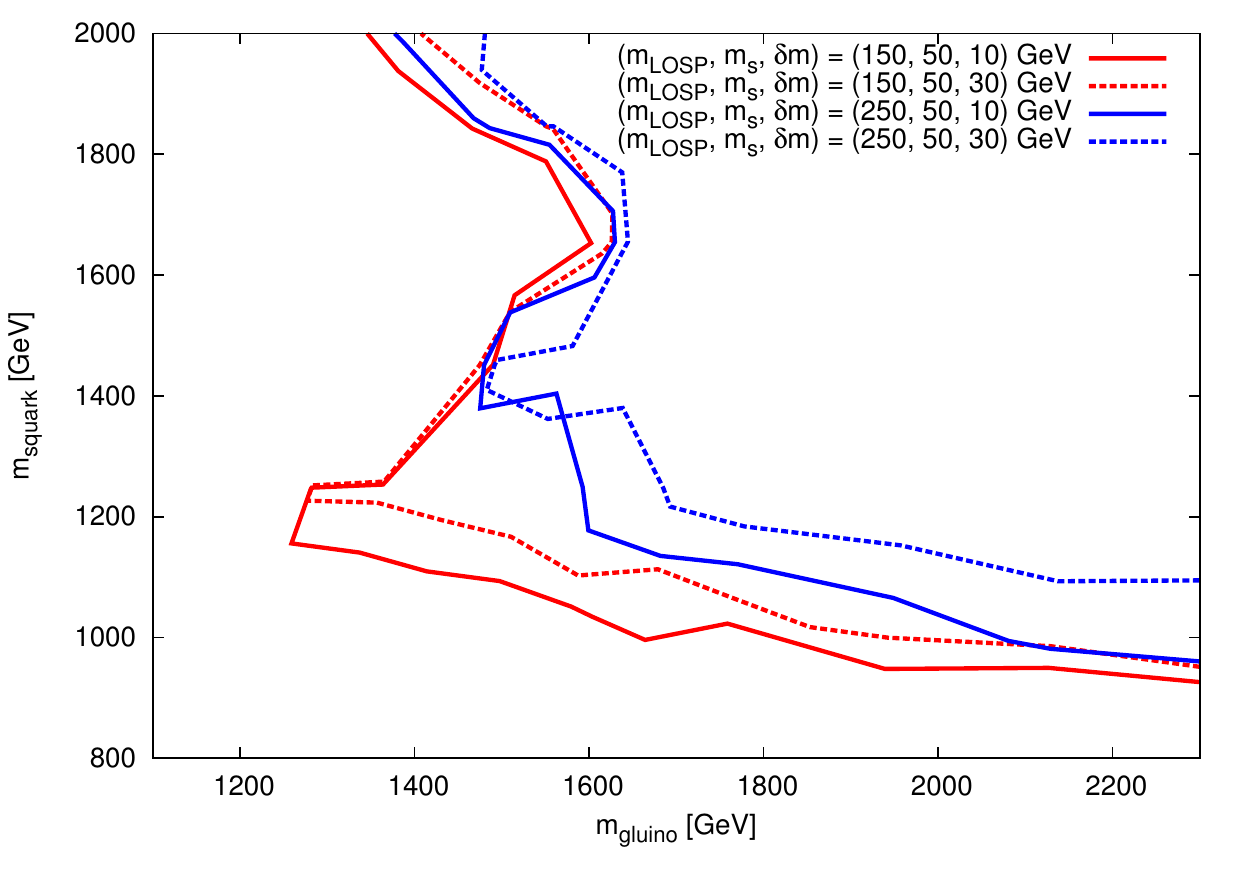}
\caption{
Constraints on the $m_{\tilde g}$-$m_{\tilde q}$ plane from an ATLAS large jet-multiplicity search \cite{Aad:2013wta}.
We take $m_{\rm LOSP}$ as 150 GeV for red lines and 250 GeV for blue lines and $\delta m$ as 10 GeV for solid lines and 30 GeV for dotted lines.
In this figure, we take $m_s$ as 50 GeV and assume the Bino and Wino are heavier than the gluino for simplicity of the analysis.
 }\label{fig:LHCRunI_constraint}
\end{figure}
In this section, we discuss the present status of our scenario.
As we have seen in Section \ref{sec:confinement}, the Hidden Valley sector contains the glueball and gluinoball supermultiplets.
For collider simulation,
it is enough to introduce a scalar boson $s$ and a fermion $\tilde s$ as well as the MSSM particles.
Here $s$ is a mixture of the glueball and gluinoball and $\tilde s$ is a mixture of their superpartners.
We also introduce the gravitino $\tilde G$ as the lightest supersymmetric particle.

\subsection{Constraints on gluino and squark masses}

The lightest supersymmetric particle in the MSSM sector is the higgsino-like neutralino $\tilde\chi_1^0$
and the mass ordering is $m_{\rm LOSP} \equiv m_{\tilde\chi_1^0} > m_{\tilde s} > m_s > m_{3/2}$.
The neutralino $\tilde\chi_1^0$ decays as $\tilde\chi_1^0 \to s \tilde s$
by using the interaction term $W_{\rm simplified} \supset \lambda_S S H_u H_d + (\kappa/3) S^3$ in \eqref{Wsimplified}.
Then $\tilde s$ decays into $s$ emitting the gravitino $\tilde G$ as discussed in Section \ref{sec:glueballino_decay}.
Finally, $s$ decays into the SM particles via mixing with the SM-like Higgs boson as discussed in Section \ref{sec:glueball_decay}.
We assume $s$, $\tilde s$, and $\tilde\chi_1^0$ decay promptly.
The gravitino would be observed as missing $E_T$ at the LHC.
However, the size of missing $E_T$ depends on the mass splitting $\delta m \equiv m_{\tilde s}-m_s$ between $\tilde s$ and $s$.
The momentum of $\tilde G$ in the rest frame of $\tilde s$ is $(m_{\tilde s}^2 - m_s^2)/2m_{\tilde s} \sim \delta m$.
A typical Lorentz boost factor in the laboratory frame is given by $m_{\tilde g, \tilde q} / m_{\tilde s}$.
Thus, the amount of missing $E_T$ is roughly $\sim (m_{\tilde g, \tilde q} / m_{\tilde s} )\delta m$
and the small mass splitting $\delta m$ suppresses the size of missing $E_T$ \cite{Fan:2011yu,Fan:2012jf}.
This mechanism weakens the constraint from null results at the LHC Run 1.

As discussed in Section \ref{sec:LOSPdecay}, the dominant decay mode for $\tilde\chi_1^0$ is $\tilde\chi_1^0 \to s \tilde s$ because of large $\kappa$.
Also, $\tilde\chi_2^0$ mostly decays into $s \tilde s$.
A subtle question is the decay of $\tilde\chi_1^\pm$.
Although strictly speaking ${\tilde \chi}^0_2$ and ${\tilde \chi}^\pm_1$ are not the LOSP,
in the higgsino multiplet these states are approximately degenerate.
Thus there is the potential for the decays ${\tilde \chi}^\pm_1 \to W^\pm {\tilde s}$ to dominate over
purely MSSM transitions like ${\tilde \chi}^\pm_1 \to W^{\pm *} {\tilde \chi}^0_1$.
The condition for decays directly to the singlino to dominate is that the transitions within the higgsino multiplet are suppressed due to small phase space, i.e.~that the mass splitting $\delta$ satisfies~\cite{Stealth3}
\be
\delta \simlt 20~{\rm GeV} \left(\frac{\lambda_S}{10^{-3}}\right)^{2/5} \left(\frac{\mu}{250~{\rm GeV}}\right)^{1/5}.
\ee
The values of $\lambda_S$ that we consider are significantly smaller than those considered in previous work on Stealth SUSY.
Nonetheless, they are typically not much smaller than $10^{-3}$. For $M_{1,2} \approx 1\, {\rm to}\, 2~{\rm TeV}$,
typical splittings among the higgsino states are $\delta \approx 2\, {\rm to}\, 5~{\rm GeV}$.
Thus, for the parameter space that we focus on we can usually assume that
$\tilde\chi_1^\pm \to W^\pm \tilde s$ is dominant for $\tilde\chi_1^\pm$.
All this assumes that there is sufficient phase space for the decays ${\tilde \chi}^\pm_1 \to W^\pm {\tilde s}$.
If these are instead decays to an off-shell $W$,
the additional phase space suppression will lead to the dominant decay being a transition within $\tilde\chi_1^\pm$ to $\tilde\chi_1^0$.
However, for now we will always consider scenarios with sufficient phase space for a two-body higgsino decay.

We show the present constraint on our models.
For simplicity, we assume the branching fraction of $\tilde s$ and higgsino multiplets $\tilde\chi_1^0$  $\tilde\chi_2^0$, $\tilde\chi_1^\pm$ as,
\begin{align}
{\rm Br}(\tilde \chi_1^0 \to s \tilde s)  = 1,\qquad
{\rm Br}(\tilde \chi_2^0 \to s \tilde s)  = 1,\qquad
{\rm Br}(\tilde \chi_1^\pm \to W^\pm \tilde s)  = 1,\qquad
{\rm Br}(\tilde s \to s \tilde G) = 1.
\end{align}
The branching fraction of $s$ can be estimated by the branching fraction of the SM Higgs boson,
which can be calculated by using \texttt{HDECAY} \cite{Djouadi:1997yw}.
For $m_s = 50~{\rm GeV}$, 
\begin{align}
{\rm Br}(s \to b\bar b) &= 0.87,\\
{\rm Br}(s \to c\bar c) &= 0.04,\\
{\rm Br}(s \to \tau\bar \tau) &= 0.07,\\
{\rm Br}(s \to gg) &= 0.02.
\end{align}
Mass dependence of the branching fractions is not significant for a small $m_s$.
We neglect other decay modes of $s$.
The decay table and the mass spectrum for other MSSM particles are calculated by \texttt{SUSYHIT} \cite{Djouadi:2006bz}.
At the LHC, SUSY particles are mainly produced by pair production of colored SUSY particles,
i.e.~$\tilde g \tilde g$, $\tilde g \tilde q$, $\tilde q \tilde q$ and $\tilde q \tilde q^*$.
In particular, in low scale gaugino mediation, squarks are mildly lighter than the gluino.
Since gluino exchange diagrams give large contributions for the squark production in such a mass spectrum,
the dominant mode of SUSY particles is $\tilde q \tilde q$ and the subdominant mode is $\tilde q \tilde q^*$.
The production cross sections for these modes are calculated at the next leading order by \texttt{Prospino 2.1} \cite{Beenakker:1996ed}.
We generate SUSY events by using \texttt{PYTHIA 8.209} \cite{Sjostrand:2014zea},
and interface them to \texttt{CheckMATE 1.2.1} \cite{Drees:2013wra} to obtain the present constraint. \texttt{CheckMATE} makes use of the DELPHES detector simulation \cite{deFavereau:2013fsa}, FastJet \cite{Cacciari:2011ma, Cacciari:2005hq}, the anti-$k_t$ jet algorithm \cite{Cacciari:2008gp}, and the $CL_s$ prescription for setting limits \cite{Read:2002hq}.
Since the dominant decay product of $\tilde\chi_1^0$ is $\tilde G + 4b$ where $b$-jets come from decays of $h$ and $s$,
the jet multiplicity in SUSY events becomes large.
Thus, we find that the most stringent bound comes from a search for large jet multiplicity with missing transverse momentum \cite{Aad:2013wta},
whose internal name in \texttt{CheckMATE} is \verb|atlas_1308_1841|.
In particular, the strongest bound is from the signal region with the number of $b$-jets larger than $2$.\footnote{We have also checked that a different ATLAS multijet search based on counting events with high jet multiplicity {\em without} a missing transverse momentum requirement \cite{TheATLAScollaboration:2013xia,Aad:2015lea} sets a somewhat weaker bound. Because this analysis is not included in \texttt{CheckMATE}, we used an independent code validated by one of the authors and discussed in \cite{Stealth3}.}
In Figure \ref{fig:LHCRunI_constraint}, we show the present constraint from the LHC Run 1 data in the $m_{\tilde g}$-$m_{\tilde q}$ plane.
We can see that the bound is weaker than the case of the MSSM mass spectrum because
smaller $\delta m$ gives the suppression of $E_T$. Thus we can see the stealth SUSY scenario works in our setup.
Let us comment on other features of this figure.
A smaller $m_{\rm LOSP}$ collimates the four $b$-jets from the decay of the LOSP.
This gives smaller jet multiplicity and reduces the efficiency of the cut, which leads to a weaker bound for a smaller $m_{\rm LOSP}$.
The constraint on the region with $m_{\tilde g} < m_{\tilde q}$ is more severe than the other side.
If $m_{\tilde g} > m_{\tilde q}$, the branching fractions of $\tilde g \to q{\tilde q}$ are almost independent of the flavor of $q$.
On the other hand, for $m_{\tilde g} < m_{\tilde q}$,
$\tilde g\to tt \chi_{1,2}^0, tb\chi_1^\pm$ become the dominant modes because we assume the LOSP is higgsino-like.
Since the top quark is a source of missing $E_T$ due to the $W$ boson, the constraint on the region with $m_{\tilde g} < m_{\tilde q}$ is more severe.

\subsection{Comments on (s)quirk phenomenology}

Let us briefly describe collider phenomenology of the new vectorlike fields.
This has been discussed in Refs.~\cite{Kang:2008ea,Burdman:2008ek,Kribs:2009fy,Harnik:2011mv}
where the fermions charged under the new gauge group are called quirks.
We mainly follow the discussions of these works and comment on some new features of the present scenario.
First, consider the scalar superpartners of quirks which we call squirks.
The electroweak doublet and color triplet squirks are pair-produced at the LHC.
Due to the soft scalar masses, these squirks are heavier than the corresponding quirk fermions.
Then, the squirk decays promptly to the quirk fermion and the hidden gaugino.
The direct pair-production rates of the SM singlet scalars are highly suppressed because their couplings to the SM particles are small.
On the other hand, collider phenomenology of the quirks is more involved.
The direct pair-production processes for the quirk fermions are given by
$p p \rightarrow \psi_f \bar{\psi}_f$,
$p p \rightarrow Z^{(*)}, \, \gamma^{(*)}  \rightarrow \psi_u^+ \bar{\psi}_d^-, \, \psi_u^0 \bar{\psi}_d^0$,
$p p \rightarrow W^{+(*)} \rightarrow \psi_u^+ \bar{\psi}_d^0$ and so on.
The heavier charged quirk fermions decay to the lighter neutral quirks.
The quirk-antiquirk pairs are joined by the hidden gauge flux strings whose
lengths are much smaller than $1 \, \rm mm$.
These bound states, the quirkonia, can lose energy via hidden glueball or gluinoball emission and radiation of many soft photons before pair annihilation.
The $\psi_f \bar{\psi}_f$ state can also radiate soft pions.
They finally annihilate in the $S$-wave states.
The dominant decays are the ones to the hidden glueballs or the gluinoballs.
As discussed above, the Hidden Valley fields decay to the SM particles,
which might lead to signals with many $b$-jets at the LHC.

The colored quirk and squirk can be produced by the gluino decay when the gluino mass is heavy enough.
In the present scenario, the heavy gluino is hardly produced at the LHC Run 1 but
is produced much more at Run 2.
In this case, we should include a possible effect on the bound on the gluino mass from the gluino decay to the colored quirk and squirk.
On the other hand, in our models, the colored quirks $\psi_f$, $\bar{\psi}_f$ preserve their own baryon number and are completely stable without any extension.
This might be incompatible with the standard cosmology if they are produced in significant numbers during reheating.
However, we can easily extend the models by adding a renormalizable operator $\Delta W = \lambda_i \bar{d} f \bar{\Psi}_i + \lambda'_i \ell \Psi_u \bar{\Psi}_i $ $(i \neq 0)$
to the superpotential and assuming the somewhat lighter singlet fermion $\bar{\psi}_i$ so that
the colored quirks can decay \cite{Martin:2010dc}.

\section{Discussions}
\label{sec:conclude}

We have proposed a framework of supersymmetric extensions of the Standard Model that can ameliorate both the SUSY Higgs mass problem and the missing superpartner problem.
New vectorlike matter fields couple to the Higgs and provide new loop contributions to its mass.
The new Yukawa couplings are sizable and large SUSY breaking is not needed to lift the Higgs mass.
To avoid a Landau pole for the new Yukawa couplings, these fields are charged under a new gauge group, which confines and leads to a Hidden Valley-like phenomenology.
Suppressing the soft masses of the new vectorlike scalars by gaugino mediation with a vanishing hidden gaugino mass
leads to an almost supersymmetric Hidden Valley sector.
Then, ordinary sparticles decay to exotic new states which decay back to Standard Model particles and gravitinos with reduced missing energy.
As a striking feature of this scenario, many $b$-jets are produced in the decay chain, in particular from decays of the Hidden Valley particles,
 and they might be observed as many displaced vertices
in jets at the LHC.
We find a viable parameter space of specific benchmark models which ameliorates both of the major phenomenological problems with supersymmetry.
At the LHC Run 2,
MSSM gluinos can be directly produced.
They partly decay to  (colored) quirks and squirks as well as the ordinary sector particles.
The produced quirks finally decay into SM particles through the Hidden Valley fields.
Since there is a mass hierarchy between the quirks and the Hidden Valley particles, 
their decays produce a parton shower and signals of supersymmetric particles have large ($b$-)jet multiplicity.
Then, it is not appropriate to use a simplified model for collider simulations.
For LHC Run 2 searches, we need to develop some techniques to deal with the parton shower of the Hidden Valley particles, possibly building on previous work done in Pythia \cite{Carloni:2010tw,Carloni:2011kk}.

We have assumed that the cutoff scale of the new Yukawa couplings and the hidden gauge coupling is relatively low compared to the usual unification scale around $10^{16} \, \rm GeV$.
This can be justified by considering multi-fold replication of the SM gauge groups.
That is, the moose (or quiver) of the SM gauge groups is spontaneously broken by some scalar link fields to the ordinary SM gauge group at some low scale.
As discussed in Ref.~\cite{ArkaniHamed:2001vr}, the successful unification of the gauge couplings is maintained and the unification scale is significantly lowered.
In addition, this model nicely accommodates gaugino mediation of SUSY breaking
\cite{Craig:2011yk}.
The SUSY-breaking source is separated from the gauge site to which the matter fields couple.
The MSSM gauge fields can couple to the source and the MSSM gauginos get nonzero masses at tree-level
while the other fields, including our new strong sector fields,
do not couple to the source and their nonzero soft masses are generated by the RG effects.
The scale where the moose of the SM gauge groups is broken corresponds to the mediation scale of SUSY breaking.
It is an interesting (and natural) alternative that the 1st and 2nd generations of quark and lepton multiplets
couple to the SUSY-breaking source.
In this case, these squarks are heavy while the 3rd generation squarks remain light so that the natural SUSY spectrum can be realized. We expect that the experimental bounds on squarks in such a scenario are significantly weaker than those we have presented in figure \ref{fig:LHCRunI_constraint}, and will resemble those discussed in \cite{Stealth3}.
The detailed analysis of this model is left for future work.

Another question in the present framework is a candidate for the dark matter in our universe.
In the usual supersymmetric models, the dark matter can be explained by the lightest neutralino or the gravitino,
depending on the scale of SUSY breaking.
In our scenario, the lightest supersymmetric particle is assumed to be the gravitino, but the correct abundance of the gravitino dark matter gives a severe constraint on the reheating temperature after the inflation.
If the gravitino is as light as $\mathcal{O} (1) \,\rm eV$, one possible candidate of the dark matter in the present model is the lightest hidden baryon.
Since there is an unbroken baryon number symmetry in our model, the lightest particle charged under the symmetry becomes stable.
It might be interesting to analyze the abundance and the observational prospect of this hidden baryon dark matter.

\section*{Acknowledgments}

We thank Ethan Neil for useful correspondence on the interpretation of lattice gauge theory calculations.
We are also grateful to Hitoshi Murayama, Yasunori Nomura, Chris Rogan and Satoshi Shirai for useful discussions.
YN is supported by a JSPS Fellowship for Research Abroad.
The work of MR is supported in part by the NSF Grant PHY-1415548.
RS is supported by a JSPS Fellowship for Young Scientists.
This work was supported in part by the National Science Foundation under Grant No. PHYS-1066293 and the hospitality of the Aspen Center for Physics.

\bibliography{ref}

\providecommand{\href}[2]{#2}\begingroup\raggedright\begin{thebibliography}{10%
0}

\bibitem{Craig:2013cxa}
N.~Craig, ``{The State of Supersymmetry after Run I of the LHC},''
\href{http://arxiv.org/abs/1309.0528}{{\ttfamily arXiv:1309.0528 [hep-ph]}}.

\bibitem{Haber:1990aw}
H.~E. Haber and R.~Hempfling, ``{Can the mass of the lightest Higgs boson of
  the minimal supersymmetric model be larger than m(Z)?},''
\href{http://dx.doi.org/10.1103/PhysRevLett.66.1815}{{\em Phys.Rev.Lett.}
  {\bfseries 66} (1991) 1815--1818}.

\bibitem{Okada:1990vk}
Y.~Okada, M.~Yamaguchi, and T.~Yanagida, ``{Upper bound of the lightest Higgs
  boson mass in the minimal supersymmetric standard model},''
\href{http://dx.doi.org/10.1143/PTP.85.1}{{\em Prog. Theor. Phys.} {\bfseries
  85} (1991) 1--6}.

\bibitem{Barbieri:1990ja}
R.~Barbieri, M.~Frigeni, and F.~Caravaglios, ``{The Supersymmetric Higgs for
  heavy superpartners},''
\href{http://dx.doi.org/10.1016/0370-2693(91)91226-L}{{\em Phys.Lett.}
  {\bfseries B258} (1991) 167--170}.

\bibitem{Draper:2011aa}
P.~Draper, P.~Meade, M.~Reece, and D.~Shih, ``{Implications of a 125 GeV Higgs
  for the MSSM and Low-Scale SUSY Breaking},''
  \href{http://dx.doi.org/10.1103/PhysRevD.85.095007}{{\em Phys.Rev.}
  {\bfseries D85} (2012) 095007},
\href{http://arxiv.org/abs/1112.3068}{{\ttfamily arXiv:1112.3068 [hep-ph]}}.

\bibitem{Arvanitaki:2012ps}
A.~Arvanitaki, N.~Craig, S.~Dimopoulos, and G.~Villadoro, ``{Mini-Split},''
  \href{http://dx.doi.org/10.1007/JHEP02(2013)126}{{\em JHEP} {\bfseries 1302}
  (2013) 126},
\href{http://arxiv.org/abs/1210.0555}{{\ttfamily arXiv:1210.0555 [hep-ph]}}.

\bibitem{Aad:2013wta}
{\bfseries ATLAS} Collaboration, G.~Aad {\em et~al.}, ``{Search for new
  phenomena in final states with large jet multiplicities and missing
  transverse momentum at sqrt(s)=8 TeV proton-proton collisions using the ATLAS
  experiment},'' \href{http://dx.doi.org/10.1007/JHEP10(2013)130}{{\em JHEP}
  {\bfseries 1310} (2013) 130},
\href{http://arxiv.org/abs/1308.1841}{{\ttfamily arXiv:1308.1841 [hep-ex]}}.

\bibitem{Aad:2014bva}
{\bfseries ATLAS} Collaboration, G.~Aad {\em et~al.}, ``{Search for direct pair
  production of the top squark in all-hadronic final states in proton-proton
  collisions at $\sqrt{s}=8$ TeV with the ATLAS detector},''
  \href{http://dx.doi.org/10.1007/JHEP09(2014)015}{{\em JHEP} {\bfseries 1409}
  (2014) 015},
\href{http://arxiv.org/abs/1406.1122}{{\ttfamily arXiv:1406.1122 [hep-ex]}}.

\bibitem{Aad:2014kra}
{\bfseries ATLAS} Collaboration, G.~Aad {\em et~al.}, ``{Search for top squark
  pair production in final states with one isolated lepton, jets, and missing
  transverse momentum in $\sqrt s =$8 TeV $pp$ collisions with the ATLAS
  detector},'' \href{http://dx.doi.org/10.1007/JHEP11(2014)118}{{\em JHEP}
  {\bfseries 1411} (2014) 118},
\href{http://arxiv.org/abs/1407.0583}{{\ttfamily arXiv:1407.0583 [hep-ex]}}.

\bibitem{Chatrchyan:2014lfa}
{\bfseries CMS} Collaboration, S.~Chatrchyan {\em et~al.}, ``{Search for new
  physics in the multijet and missing transverse momentum final state in
  proton-proton collisions at $\sqrt{s}$ = 8 TeV},''
\href{http://arxiv.org/abs/1402.4770}{{\ttfamily arXiv:1402.4770 [hep-ex]}}.

\bibitem{Khachatryan:2014doa}
{\bfseries CMS} Collaboration, V.~Khachatryan {\em et~al.}, ``{Search for
  top-squark pairs decaying into Higgs or Z bosons in pp collisions at
  $\sqrt{s}$ = 8 TeV},''
\href{http://arxiv.org/abs/1405.3886}{{\ttfamily arXiv:1405.3886 [hep-ex]}}.

\bibitem{Khachatryan:2015wza}
{\bfseries CMS} Collaboration, V.~Khachatryan {\em et~al.}, ``{Searches for
  third generation squark production in fully hadronic final states in
  proton-proton collisions at sqrt(s)=8 TeV},''
\href{http://arxiv.org/abs/1503.08037}{{\ttfamily arXiv:1503.08037 [hep-ex]}}.

\bibitem{Drees:1987tp}
M.~Drees, ``{Comment on `Higgs Boson Mass Bound in $E(6)$ Based Supersymmetric
  Theories.'},''
\href{http://dx.doi.org/10.1103/PhysRevD.35.2910}{{\em Phys.Rev.} {\bfseries
  D35} (1987) 2910--2913}.

\bibitem{Drees:1988fc}
M.~Drees, ``{Supersymmetric Models with Extended Higgs Sector},''
\href{http://dx.doi.org/10.1142/S0217751X89001448}{{\em Int.J.Mod.Phys.}
  {\bfseries A4} (1989) 3635}.

\bibitem{Espinosa:1991gr}
J.~Espinosa and M.~Quiros, ``{On Higgs boson masses in nonminimal
  supersymmetric standard models},''
\href{http://dx.doi.org/10.1016/0370-2693(92)91846-2}{{\em Phys.Lett.}
  {\bfseries B279} (1992) 92--97}.

\bibitem{Randall:2002talk}
L.~Randall, ``{talk at The 10th International Conference on Supersymmetry and
  Unification of Fundamental Interactions (SUSY02), Hamburg, Germany, 17-23 Jun
  2002},'' {\em unpublished} (2002) .

\bibitem{Batra:2003nj}
P.~Batra, A.~Delgado, D.~E. Kaplan, and T.~M. Tait, ``{The Higgs mass bound in
  gauge extensions of the minimal supersymmetric standard model},''
  \href{http://dx.doi.org/10.1088/1126-6708/2004/02/043}{{\em JHEP} {\bfseries
  0402} (2004) 043},
\href{http://arxiv.org/abs/hep-ph/0309149}{{\ttfamily arXiv:hep-ph/0309149
  [hep-ph]}}.

\bibitem{Casas:2003jx}
J.~Casas, J.~Espinosa, and I.~Hidalgo, ``{The MSSM fine tuning problem: A Way
  out},'' \href{http://dx.doi.org/10.1088/1126-6708/2004/01/008}{{\em JHEP}
  {\bfseries 0401} (2004) 008},
\href{http://arxiv.org/abs/hep-ph/0310137}{{\ttfamily arXiv:hep-ph/0310137
  [hep-ph]}}.

\bibitem{Barbieri:2006bg}
R.~Barbieri, L.~J. Hall, Y.~Nomura, and V.~S. Rychkov, ``{Supersymmetry without
  a Light Higgs Boson},''
  \href{http://dx.doi.org/10.1103/PhysRevD.75.035007}{{\em Phys.Rev.}
  {\bfseries D75} (2007) 035007},
\href{http://arxiv.org/abs/hep-ph/0607332}{{\ttfamily arXiv:hep-ph/0607332
  [hep-ph]}}.

\bibitem{Dine:2007xi}
M.~Dine, N.~Seiberg, and S.~Thomas, ``{Higgs physics as a window beyond the
  MSSM (BMSSM)},'' \href{http://dx.doi.org/10.1103/PhysRevD.76.095004}{{\em
  Phys.Rev.} {\bfseries D76} (2007) 095004},
\href{http://arxiv.org/abs/0707.0005}{{\ttfamily arXiv:0707.0005 [hep-ph]}}.

\bibitem{Evans:2011bea}
J.~L. Evans, M.~Ibe, and T.~T. Yanagida, ``{Relatively Heavy Higgs Boson in
  More Generic Gauge Mediation},''
  \href{http://dx.doi.org/10.1016/j.physletb.2011.10.031}{{\em Phys. Lett.}
  {\bfseries B705} (2011) 342--348},
\href{http://arxiv.org/abs/1107.3006}{{\ttfamily arXiv:1107.3006 [hep-ph]}}.

\bibitem{Kang:2012ra}
Z.~Kang, T.~Li, T.~Liu, C.~Tong, and J.~M. Yang, ``{A Heavy SM-like Higgs and a
  Light Stop from Yukawa-Deflected Gauge Mediation},''
  \href{http://dx.doi.org/10.1103/PhysRevD.86.095020}{{\em Phys. Rev.}
  {\bfseries D86} (2012) 095020},
\href{http://arxiv.org/abs/1203.2336}{{\ttfamily arXiv:1203.2336 [hep-ph]}}.

\bibitem{Craig:2012xp}
N.~Craig, S.~Knapen, D.~Shih, and Y.~Zhao, ``{A Complete Model of Low-Scale
  Gauge Mediation},'' \href{http://dx.doi.org/10.1007/JHEP03(2013)154}{{\em
  JHEP} {\bfseries 03} (2013) 154},
\href{http://arxiv.org/abs/1206.4086}{{\ttfamily arXiv:1206.4086 [hep-ph]}}.

\bibitem{Abdullah:2012tq}
M.~Abdullah, I.~Galon, Y.~Shadmi, and Y.~Shirman, ``{Flavored Gauge Mediation,
  A Heavy Higgs, and Supersymmetric Alignment},''
  \href{http://dx.doi.org/10.1007/JHEP06(2013)057}{{\em JHEP} {\bfseries 06}
  (2013) 057},
\href{http://arxiv.org/abs/1209.4904}{{\ttfamily arXiv:1209.4904 [hep-ph]}}.

\bibitem{Evans:2013kxa}
J.~A. Evans and D.~Shih, ``{Surveying Extended GMSB Models with $m$$_{h}$=125
  GeV},'' \href{http://dx.doi.org/10.1007/JHEP08(2013)093}{{\em JHEP}
  {\bfseries 08} (2013) 093},
\href{http://arxiv.org/abs/1303.0228}{{\ttfamily arXiv:1303.0228 [hep-ph]}}.

\bibitem{Basirnia:2015vga}
A.~Basirnia, D.~Egana-Ugrinovic, S.~Knapen, and D.~Shih, ``{125 GeV Higgs from
  Tree-Level $A$-terms},''
  \href{http://dx.doi.org/10.1007/JHEP06(2015)144}{{\em JHEP} {\bfseries 06}
  (2015) 144},
\href{http://arxiv.org/abs/1501.00997}{{\ttfamily arXiv:1501.00997 [hep-ph]}}.

\bibitem{Knapen:2015qba}
S.~Knapen, D.~Redigolo, and D.~Shih, ``{General Gauge Mediation at the Weak
  Scale},''
\href{http://arxiv.org/abs/1507.04364}{{\ttfamily arXiv:1507.04364 [hep-ph]}}.

\bibitem{Moroi:1991mg}
T.~Moroi and Y.~Okada, ``{Radiative corrections to Higgs masses in the
  supersymmetric model with an extra family and antifamily},''
\href{http://dx.doi.org/10.1142/S0217732392000124}{{\em Mod. Phys. Lett.}
  {\bfseries A7} (1992) 187--200}.

\bibitem{Moroi:1992zk}
T.~Moroi and Y.~Okada, ``{Upper bound of the lightest neutral Higgs mass in
  extended supersymmetric Standard Models},''
\href{http://dx.doi.org/10.1016/0370-2693(92)90091-H}{{\em Phys. Lett.}
  {\bfseries B295} (1992) 73--78}.

\bibitem{Babu:2004xg}
K.~Babu, I.~Gogoladze, and C.~Kolda, ``{Perturbative unification and Higgs
  boson mass bounds},''
\href{http://arxiv.org/abs/hep-ph/0410085}{{\ttfamily arXiv:hep-ph/0410085
  [hep-ph]}}.

\bibitem{Babu:2008ge}
K.~Babu, I.~Gogoladze, M.~U. Rehman, and Q.~Shafi, ``{Higgs Boson Mass,
  Sparticle Spectrum and Little Hierarchy Problem in Extended MSSM},''
  \href{http://dx.doi.org/10.1103/PhysRevD.78.055017}{{\em Phys.Rev.}
  {\bfseries D78} (2008) 055017},
\href{http://arxiv.org/abs/0807.3055}{{\ttfamily arXiv:0807.3055 [hep-ph]}}.

\bibitem{Martin:2009bg}
S.~P. Martin, ``{Extra vector-like matter and the lightest Higgs scalar boson
  mass in low-energy supersymmetry},''
  \href{http://dx.doi.org/10.1103/PhysRevD.81.035004}{{\em Phys.Rev.}
  {\bfseries D81} (2010) 035004},
\href{http://arxiv.org/abs/0910.2732}{{\ttfamily arXiv:0910.2732 [hep-ph]}}.

\bibitem{Graham:2009gy}
P.~W. Graham, A.~Ismail, S.~Rajendran, and P.~Saraswat, ``{A Little Solution to
  the Little Hierarchy Problem: A Vector-like Generation},''
  \href{http://dx.doi.org/10.1103/PhysRevD.81.055016}{{\em Phys.Rev.}
  {\bfseries D81} (2010) 055016},
\href{http://arxiv.org/abs/0910.3020}{{\ttfamily arXiv:0910.3020 [hep-ph]}}.

\bibitem{Martin:2010dc}
S.~P. Martin, ``{Raising the Higgs mass with Yukawa couplings for isotriplets
  in vector-like extensions of minimal supersymmetry},''
  \href{http://dx.doi.org/10.1103/PhysRevD.82.055019}{{\em Phys.Rev.}
  {\bfseries D82} (2010) 055019},
\href{http://arxiv.org/abs/1006.4186}{{\ttfamily arXiv:1006.4186 [hep-ph]}}.

\bibitem{Martin:2010kk}
S.~P. Martin, ``{Quirks in supersymmetry with gauge coupling unification},''
  \href{http://dx.doi.org/10.1103/PhysRevD.83.035019}{{\em Phys.Rev.}
  {\bfseries D83} (2011) 035019},
\href{http://arxiv.org/abs/1012.2072}{{\ttfamily arXiv:1012.2072 [hep-ph]}}.

\bibitem{Asano:2011zt}
M.~Asano, T.~Moroi, R.~Sato, and T.~T. Yanagida, ``{Non-anomalous Discrete
  R-symmetry, Extra Matters, and Enhancement of the Lightest SUSY Higgs
  Mass},'' \href{http://dx.doi.org/10.1016/j.physletb.2011.10.025}{{\em
  Phys.Lett.} {\bfseries B705} (2011) 337--341},
\href{http://arxiv.org/abs/1108.2402}{{\ttfamily arXiv:1108.2402 [hep-ph]}}.

\bibitem{Endo:2011mc}
M.~Endo, K.~Hamaguchi, S.~Iwamoto, and N.~Yokozaki, ``{Higgs Mass and Muon
  Anomalous Magnetic Moment in Supersymmetric Models with Vector-Like
  Matters},'' \href{http://dx.doi.org/10.1103/PhysRevD.84.075017}{{\em Phys.
  Rev.} {\bfseries D84} (2011) 075017},
\href{http://arxiv.org/abs/1108.3071}{{\ttfamily arXiv:1108.3071 [hep-ph]}}.

\bibitem{Heckman:2011bb}
J.~J. Heckman, P.~Kumar, C.~Vafa, and B.~Wecht, ``{Electroweak Symmetry
  Breaking in the DSSM},''
  \href{http://dx.doi.org/10.1007/JHEP01(2012)156}{{\em JHEP} {\bfseries 01}
  (2012) 156},
\href{http://arxiv.org/abs/1108.3849}{{\ttfamily arXiv:1108.3849 [hep-ph]}}.

\bibitem{Moroi:2011aa}
T.~Moroi, R.~Sato, and T.~T. Yanagida, ``{Extra Matters Decree the Relatively
  Heavy Higgs of Mass about 125 GeV in the Supersymmetric Model},''
  \href{http://dx.doi.org/10.1016/j.physletb.2012.02.012}{{\em Phys.Lett.}
  {\bfseries B709} (2012) 218--221},
\href{http://arxiv.org/abs/1112.3142}{{\ttfamily arXiv:1112.3142 [hep-ph]}}.

\bibitem{Endo:2011xq}
M.~Endo, K.~Hamaguchi, S.~Iwamoto, and N.~Yokozaki, ``{Higgs mass, muon g-2,
  and LHC prospects in gauge mediation models with vector-like matters},''
  \href{http://dx.doi.org/10.1103/PhysRevD.85.095012}{{\em Phys. Rev.}
  {\bfseries D85} (2012) 095012},
\href{http://arxiv.org/abs/1112.5653}{{\ttfamily arXiv:1112.5653 [hep-ph]}}.

\bibitem{Endo:2012rd}
M.~Endo, K.~Hamaguchi, S.~Iwamoto, and N.~Yokozaki, ``{Vacuum Stability Bound
  on Extended GMSB Models},''
  \href{http://dx.doi.org/10.1007/JHEP06(2012)060}{{\em JHEP} {\bfseries 06}
  (2012) 060},
\href{http://arxiv.org/abs/1202.2751}{{\ttfamily arXiv:1202.2751 [hep-ph]}}.

\bibitem{Evans:2012uf}
J.~L. Evans, M.~Ibe, and T.~T. Yanagida, ``{The Lightest Higgs Boson Mass in
  the MSSM with Strongly Interacting Spectators},''
  \href{http://dx.doi.org/10.1103/PhysRevD.86.015017}{{\em Phys. Rev.}
  {\bfseries D86} (2012) 015017},
\href{http://arxiv.org/abs/1204.6085}{{\ttfamily arXiv:1204.6085 [hep-ph]}}.

\bibitem{Martin:2012dg}
S.~P. Martin and J.~D. Wells, ``{Implications of gauge-mediated supersymmetry
  breaking with vector-like quarks and a ~125 GeV Higgs boson},''
  \href{http://dx.doi.org/10.1103/PhysRevD.86.035017}{{\em Phys.Rev.}
  {\bfseries D86} (2012) 035017},
\href{http://arxiv.org/abs/1206.2956}{{\ttfamily arXiv:1206.2956 [hep-ph]}}.

\bibitem{Kitano:2012wv}
R.~Kitano, M.~A. Luty, and Y.~Nakai, ``{Partially Composite Higgs in
  Supersymmetry},'' \href{http://dx.doi.org/10.1007/JHEP08(2012)111}{{\em JHEP}
  {\bfseries 08} (2012) 111},
\href{http://arxiv.org/abs/1206.4053}{{\ttfamily arXiv:1206.4053 [hep-ph]}}.

\bibitem{Endo:2012cc}
M.~Endo, K.~Hamaguchi, K.~Ishikawa, S.~Iwamoto, and N.~Yokozaki, ``{Gauge
  Mediation Models with Vectorlike Matters at the LHC},''
  \href{http://dx.doi.org/10.1007/JHEP01(2013)181}{{\em JHEP} {\bfseries 01}
  (2013) 181},
\href{http://arxiv.org/abs/1212.3935}{{\ttfamily arXiv:1212.3935 [hep-ph]}}.

\bibitem{Lalak:2015xea}
Z.~Lalak, M.~Lewicki, and J.~D. Wells, ``{Higgs boson mass and high-luminosity
  LHC probes of supersymmetry with vectorlike top quark},''
  \href{http://dx.doi.org/10.1103/PhysRevD.91.095022}{{\em Phys.Rev.}
  {\bfseries D91} no.~9, (2015) 095022},
\href{http://arxiv.org/abs/1502.05702}{{\ttfamily arXiv:1502.05702 [hep-ph]}}.

\bibitem{Barbier:2004ez}
R.~Barbier, C.~Berat, M.~Besancon, M.~Chemtob, A.~Deandrea, E.~Dudas, P.~Fayet,
  S.~Lavignac, G.~Moreau, E.~Perez, and Y.~Sirois, ``{R-parity violating
  supersymmetry},'' \href{http://dx.doi.org/10.1016/j.physrep.2005.08.006}{{\em
  Phys.Rept.} {\bfseries 420} (2005) 1--202},
\href{http://arxiv.org/abs/hep-ph/0406039}{{\ttfamily arXiv:hep-ph/0406039
  [hep-ph]}}.

\bibitem{Csaki:2011ge}
C.~Csaki, Y.~Grossman, and B.~Heidenreich, ``{MFV SUSY: A Natural Theory for
  R-Parity Violation},''
  \href{http://dx.doi.org/10.1103/PhysRevD.85.095009}{{\em Phys.Rev.}
  {\bfseries D85} (2012) 095009},
\href{http://arxiv.org/abs/1111.1239}{{\ttfamily arXiv:1111.1239 [hep-ph]}}.

\bibitem{Graham:2012th}
P.~W. Graham, D.~E. Kaplan, S.~Rajendran, and P.~Saraswat, ``{Displaced
  Supersymmetry},'' \href{http://dx.doi.org/10.1007/JHEP07(2012)149}{{\em JHEP}
  {\bfseries 1207} (2012) 149},
\href{http://arxiv.org/abs/1204.6038}{{\ttfamily arXiv:1204.6038 [hep-ph]}}.

\bibitem{Csaki:2013jza}
C.~Csaki, E.~Kuflik, and T.~Volansky, ``{Dynamical R-Parity Violation},''
  \href{http://dx.doi.org/10.1103/PhysRevLett.112.131801}{{\em Phys.Rev.Lett.}
  {\bfseries 112} (2014) 131801},
\href{http://arxiv.org/abs/1309.5957}{{\ttfamily arXiv:1309.5957 [hep-ph]}}.

\bibitem{Graham:2014vya}
P.~W. Graham, S.~Rajendran, and P.~Saraswat, ``{Supersymmetric crevices:
  Missing signatures of R -parity violation at the LHC},''
  \href{http://dx.doi.org/10.1103/PhysRevD.90.075005}{{\em Phys.Rev.}
  {\bfseries D90} no.~7, (2014) 075005},
\href{http://arxiv.org/abs/1403.7197}{{\ttfamily arXiv:1403.7197 [hep-ph]}}.

\bibitem{Heidenreich:2014jpa}
B.~Heidenreich and Y.~Nakai, ``{Natural Supersymmetry in Warped Space},''
  \href{http://dx.doi.org/10.1007/JHEP10(2014)182}{{\em JHEP} {\bfseries 1410}
  (2014) 182},
\href{http://arxiv.org/abs/1407.5095}{{\ttfamily arXiv:1407.5095 [hep-ph]}}.

\bibitem{Strassler:2006im}
M.~J. Strassler and K.~M. Zurek, ``{Echoes of a hidden valley at hadron
  colliders},'' \href{http://dx.doi.org/10.1016/j.physletb.2007.06.055}{{\em
  Phys.Lett.} {\bfseries B651} (2007) 374--379},
\href{http://arxiv.org/abs/hep-ph/0604261}{{\ttfamily arXiv:hep-ph/0604261
  [hep-ph]}}.

\bibitem{Strassler:2006qa}
M.~J. Strassler, ``{Possible effects of a hidden valley on supersymmetric
  phenomenology},''
\href{http://arxiv.org/abs/hep-ph/0607160}{{\ttfamily arXiv:hep-ph/0607160
  [hep-ph]}}.

\bibitem{Fan:2011yu}
J.~Fan, M.~Reece, and J.~T. Ruderman, ``{Stealth Supersymmetry},''
  \href{http://dx.doi.org/10.1007/JHEP11(2011)012}{{\em JHEP} {\bfseries 1111}
  (2011) 012},
\href{http://arxiv.org/abs/1105.5135}{{\ttfamily arXiv:1105.5135 [hep-ph]}}.

\bibitem{Fan:2012jf}
J.~Fan, M.~Reece, and J.~T. Ruderman, ``{A Stealth Supersymmetry Sampler},''
  \href{http://dx.doi.org/10.1007/JHEP07(2012)196}{{\em JHEP} {\bfseries 1207}
  (2012) 196},
\href{http://arxiv.org/abs/1201.4875}{{\ttfamily arXiv:1201.4875 [hep-ph]}}.

\bibitem{Stealth3}
J.~Fan, R.~Krall, D.~Pinner, M.~Reece, and J.~T. Ruderman, ``{Natural Stealth
  SUSY Simplified Models (to appear)},''
\href{http://arxiv.org/abs/15mm.nnnnn}{{\ttfamily arXiv:15mm.nnnnn [hep-ph]}}.

\bibitem{Martin:2007gf}
S.~P. Martin, ``{Compressed supersymmetry and natural neutralino dark matter
  from top squark-mediated annihilation to top quarks},''
  \href{http://dx.doi.org/10.1103/PhysRevD.75.115005}{{\em Phys.Rev.}
  {\bfseries D75} (2007) 115005},
\href{http://arxiv.org/abs/hep-ph/0703097}{{\ttfamily arXiv:hep-ph/0703097
  [hep-ph]}}.

\bibitem{Baer:2007uz}
H.~Baer, A.~Box, E.-K. Park, and X.~Tata, ``{Implications of compressed
  supersymmetry for collider and dark matter searches},''
  \href{http://dx.doi.org/10.1088/1126-6708/2007/08/060}{{\em JHEP} {\bfseries
  0708} (2007) 060},
\href{http://arxiv.org/abs/0707.0618}{{\ttfamily arXiv:0707.0618 [hep-ph]}}.

\bibitem{Martin:2008aw}
S.~P. Martin, ``{Exploring compressed supersymmetry with same-sign top quarks
  at the Large Hadron Collider},''
  \href{http://dx.doi.org/10.1103/PhysRevD.78.055019}{{\em Phys.Rev.}
  {\bfseries D78} (2008) 055019},
\href{http://arxiv.org/abs/0807.2820}{{\ttfamily arXiv:0807.2820 [hep-ph]}}.

\bibitem{LeCompte:2011fh}
T.~J. LeCompte and S.~P. Martin, ``{Compressed supersymmetry after 1/fb at the
  Large Hadron Collider},''
  \href{http://dx.doi.org/10.1103/PhysRevD.85.035023}{{\em Phys.Rev.}
  {\bfseries D85} (2012) 035023},
\href{http://arxiv.org/abs/1111.6897}{{\ttfamily arXiv:1111.6897 [hep-ph]}}.

\bibitem{Dreiner:2012gx}
H.~K. Dreiner, M.~Kramer, and J.~Tattersall, ``{How low can SUSY go? Matching,
  monojets and compressed spectra},''
  \href{http://dx.doi.org/10.1209/0295-5075/99/61001}{{\em Europhys.Lett.}
  {\bfseries 99} (2012) 61001},
\href{http://arxiv.org/abs/1207.1613}{{\ttfamily arXiv:1207.1613 [hep-ph]}}.

\bibitem{Dutta:2013gga}
B.~Dutta, W.~Flanagan, A.~Gurrola, W.~Johns, T.~Kamon, P.~Sheldon, K.~Sinha,
  K.~Wang, and S.~Wu, ``{Probing compressed top squark scenarios at the LHC at
  14 TeV},'' \href{http://dx.doi.org/10.1103/PhysRevD.90.095022}{{\em Phys.
  Rev.} {\bfseries D90} no.~9, (2014) 095022},
\href{http://arxiv.org/abs/1312.1348}{{\ttfamily arXiv:1312.1348 [hep-ph]}}.

\bibitem{Bhattacherjee:2013wna}
B.~Bhattacherjee, A.~Choudhury, K.~Ghosh, and S.~Poddar, ``{Compressed
  supersymmetry at 14 TeV LHC},''
  \href{http://dx.doi.org/10.1103/PhysRevD.89.037702}{{\em Phys.Rev.}
  {\bfseries D89} no.~3, (2014) 037702},
\href{http://arxiv.org/abs/1308.1526}{{\ttfamily arXiv:1308.1526 [hep-ph]}}.

\bibitem{Fox:2002bu}
P.~J. Fox, A.~E. Nelson, and N.~Weiner, ``{Dirac gaugino masses and supersoft
  supersymmetry breaking},''
  \href{http://dx.doi.org/10.1088/1126-6708/2002/08/035}{{\em JHEP} {\bfseries
  0208} (2002) 035},
\href{http://arxiv.org/abs/hep-ph/0206096}{{\ttfamily arXiv:hep-ph/0206096
  [hep-ph]}}.

\bibitem{Nelson:2002ca}
A.~E. Nelson, N.~Rius, V.~Sanz, and M.~Unsal, ``{The Minimal supersymmetric
  model without a mu term},''
  \href{http://dx.doi.org/10.1088/1126-6708/2002/08/039}{{\em JHEP} {\bfseries
  0208} (2002) 039},
\href{http://arxiv.org/abs/hep-ph/0206102}{{\ttfamily arXiv:hep-ph/0206102
  [hep-ph]}}.

\bibitem{Kribs:2007ac}
G.~D. Kribs, E.~Poppitz, and N.~Weiner, ``{Flavor in supersymmetry with an
  extended R-symmetry},''
  \href{http://dx.doi.org/10.1103/PhysRevD.78.055010}{{\em Phys.Rev.}
  {\bfseries D78} (2008) 055010},
\href{http://arxiv.org/abs/0712.2039}{{\ttfamily arXiv:0712.2039 [hep-ph]}}.

\bibitem{Kribs:2012gx}
G.~D. Kribs and A.~Martin, ``{Supersoft Supersymmetry is Super-Safe},''
  \href{http://dx.doi.org/10.1103/PhysRevD.85.115014}{{\em Phys.Rev.}
  {\bfseries D85} (2012) 115014},
\href{http://arxiv.org/abs/1203.4821}{{\ttfamily arXiv:1203.4821 [hep-ph]}}.

\bibitem{Alves:2015kia}
D.~S.~M. Alves, J.~Galloway, M.~McCullough, and N.~Weiner, ``{Goldstone
  Gauginos},'' \href{http://dx.doi.org/10.1103/PhysRevLett.115.161801}{{\em
  Phys. Rev. Lett.} {\bfseries 115} no.~16, (2015) 161801},
\href{http://arxiv.org/abs/1502.03819}{{\ttfamily arXiv:1502.03819 [hep-ph]}}.

\bibitem{Alves:2013wra}
D.~S.~M. Alves, J.~Liu, and N.~Weiner, ``{Hiding Missing Energy in Missing
  Energy},'' \href{http://dx.doi.org/10.1007/JHEP04(2015)088}{{\em JHEP}
  {\bfseries 1504} (2015) 088},
\href{http://arxiv.org/abs/1312.4965}{{\ttfamily arXiv:1312.4965 [hep-ph]}}.

\bibitem{Kyae:2013hda}
B.~Kyae and C.~S. Shin, ``{Vector-like leptons and extra gauge symmetry for the
  natural Higgs boson},'' \href{http://dx.doi.org/10.1007/JHEP06(2013)102}{{\em
  JHEP} {\bfseries 1306} (2013) 102},
\href{http://arxiv.org/abs/1303.6703}{{\ttfamily arXiv:1303.6703 [hep-ph]}}.

\bibitem{Giudice:1998bp}
G.~F. Giudice and R.~Rattazzi, ``{Theories with gauge mediated supersymmetry
  breaking},'' \href{http://dx.doi.org/10.1016/S0370-1573(99)00042-3}{{\em
  Phys. Rept.} {\bfseries 322} (1999) 419--499},
\href{http://arxiv.org/abs/hep-ph/9801271}{{\ttfamily arXiv:hep-ph/9801271
  [hep-ph]}}.

\bibitem{Kaplan:1999ac}
D.~E. Kaplan, G.~D. Kribs, and M.~Schmaltz, ``{Supersymmetry breaking through
  transparent extra dimensions},''
  \href{http://dx.doi.org/10.1103/PhysRevD.62.035010}{{\em Phys. Rev.}
  {\bfseries D62} (2000) 035010},
\href{http://arxiv.org/abs/hep-ph/9911293}{{\ttfamily arXiv:hep-ph/9911293
  [hep-ph]}}.

\bibitem{Chacko:1999mi}
Z.~Chacko, M.~A. Luty, A.~E. Nelson, and E.~Ponton, ``{Gaugino mediated
  supersymmetry breaking},''
  \href{http://dx.doi.org/10.1088/1126-6708/2000/01/003}{{\em JHEP} {\bfseries
  01} (2000) 003},
\href{http://arxiv.org/abs/hep-ph/9911323}{{\ttfamily arXiv:hep-ph/9911323
  [hep-ph]}}.

\bibitem{Han:2007ae}
T.~Han, Z.~Si, K.~M. Zurek, and M.~J. Strassler, ``{Phenomenology of hidden
  valleys at hadron colliders},''
  \href{http://dx.doi.org/10.1088/1126-6708/2008/07/008}{{\em JHEP} {\bfseries
  0807} (2008) 008},
\href{http://arxiv.org/abs/0712.2041}{{\ttfamily arXiv:0712.2041 [hep-ph]}}.

\bibitem{Strassler:2008bv}
M.~J. Strassler, ``{Why Unparticle Models with Mass Gaps are Examples of Hidden
  Valleys},''
\href{http://arxiv.org/abs/0801.0629}{{\ttfamily arXiv:0801.0629 [hep-ph]}}.

\bibitem{Zurek:2010xf}
K.~M. Zurek, ``{TASI 2009 Lectures: Searching for Unexpected Physics at the
  LHC},''
\href{http://arxiv.org/abs/1001.2563}{{\ttfamily arXiv:1001.2563 [hep-ph]}}.

\bibitem{Kang:2008ea}
J.~Kang and M.~A. Luty, ``{Macroscopic Strings and 'Quirks' at Colliders},''
  \href{http://dx.doi.org/10.1088/1126-6708/2009/11/065}{{\em JHEP} {\bfseries
  0911} (2009) 065},
\href{http://arxiv.org/abs/0805.4642}{{\ttfamily arXiv:0805.4642 [hep-ph]}}.

\bibitem{Burdman:2008ek}
G.~Burdman, Z.~Chacko, H.-S. Goh, R.~Harnik, and C.~A. Krenke, ``{The Quirky
  Collider Signals of Folded Supersymmetry},''
  \href{http://dx.doi.org/10.1103/PhysRevD.78.075028}{{\em Phys.Rev.}
  {\bfseries D78} (2008) 075028},
\href{http://arxiv.org/abs/0805.4667}{{\ttfamily arXiv:0805.4667 [hep-ph]}}.

\bibitem{Kribs:2009fy}
G.~D. Kribs, T.~S. Roy, J.~Terning, and K.~M. Zurek, ``{Quirky Composite Dark
  Matter},'' \href{http://dx.doi.org/10.1103/PhysRevD.81.095001}{{\em
  Phys.Rev.} {\bfseries D81} (2010) 095001},
\href{http://arxiv.org/abs/0909.2034}{{\ttfamily arXiv:0909.2034 [hep-ph]}}.

\bibitem{Harnik:2011mv}
R.~Harnik, G.~D. Kribs, and A.~Martin, ``{Quirks at the Tevatron and Beyond},''
  \href{http://dx.doi.org/10.1103/PhysRevD.84.035029}{{\em Phys.Rev.}
  {\bfseries D84} (2011) 035029},
\href{http://arxiv.org/abs/1106.2569}{{\ttfamily arXiv:1106.2569 [hep-ph]}}.

\bibitem{Brignole:2000kg}
A.~Brignole, ``{One loop Kahler potential in non renormalizable theories},''
  \href{http://dx.doi.org/10.1016/S0550-3213(00)00211-X}{{\em Nucl.Phys.}
  {\bfseries B579} (2000) 101--116},
\href{http://arxiv.org/abs/hep-th/0001121}{{\ttfamily arXiv:hep-th/0001121
  [hep-th]}}.

\bibitem{Goodsell:2014bna}
M.~D. Goodsell, K.~Nickel, and F.~Staub, ``{Two-Loop Higgs mass calculations in
  supersymmetric models beyond the MSSM with SARAH and SPheno},''
  \href{http://dx.doi.org/10.1140/epjc/s10052-014-3247-y}{{\em Eur. Phys. J.}
  {\bfseries C75} no.~1, (2015) 32},
\href{http://arxiv.org/abs/1411.0675}{{\ttfamily arXiv:1411.0675 [hep-ph]}}.

\bibitem{Goodsell:2015ira}
M.~Goodsell, K.~Nickel, and F.~Staub, ``{Generic two-loop Higgs mass
  calculation from a diagrammatic approach},''
  \href{http://dx.doi.org/10.1140/epjc/s10052-015-3494-6}{{\em Eur. Phys. J.}
  {\bfseries C75} no.~6, (2015) 290},
\href{http://arxiv.org/abs/1503.03098}{{\ttfamily arXiv:1503.03098 [hep-ph]}}.

\bibitem{Nickel:2015dna}
K.~Nickel and F.~Staub, ``{Precise determination of the Higgs mass in
  supersymmetric models with vectorlike tops and the impact on naturalness in
  minimal GMSB},'' \href{http://dx.doi.org/10.1007/JHEP07(2015)139}{{\em JHEP}
  {\bfseries 07} (2015) 139},
\href{http://arxiv.org/abs/1505.06077}{{\ttfamily arXiv:1505.06077 [hep-ph]}}.

\bibitem{Staub:2008uz}
F.~Staub, ``{SARAH},''
\href{http://arxiv.org/abs/0806.0538}{{\ttfamily arXiv:0806.0538 [hep-ph]}}.

\bibitem{Staub:2013tta}
F.~Staub, ``{SARAH 4: A tool for (not only SUSY) model builders},''
  \href{http://dx.doi.org/10.1016/j.cpc.2014.02.018}{{\em Comput.Phys.Commun.}
  {\bfseries 185} (2014) 1773--1790},
\href{http://arxiv.org/abs/1309.7223}{{\ttfamily arXiv:1309.7223 [hep-ph]}}.

\bibitem{Chetyrkin:1997sg}
K.~G. Chetyrkin, B.~A. Kniehl, and M.~Steinhauser, ``{Strong coupling constant
  with flavor thresholds at four loops in the MS scheme},''
  \href{http://dx.doi.org/10.1103/PhysRevLett.79.2184}{{\em Phys. Rev. Lett.}
  {\bfseries 79} (1997) 2184--2187},
\href{http://arxiv.org/abs/hep-ph/9706430}{{\ttfamily arXiv:hep-ph/9706430
  [hep-ph]}}.

\bibitem{Kim:2015dpa}
J.~S. Kim, D.~Schmeier, and J.~Tattersall, ``{Naughty or Nice? The Role of the
  `N' in the Natural NMSSM for the LHC},''
\href{http://arxiv.org/abs/1510.04871}{{\ttfamily arXiv:1510.04871 [hep-ph]}}.

\bibitem{Cohen:1996vb}
A.~G. Cohen, D.~Kaplan, and A.~Nelson, ``{The More minimal supersymmetric
  standard model},''
  \href{http://dx.doi.org/10.1016/S0370-2693(96)01183-5}{{\em Phys.Lett.}
  {\bfseries B388} (1996) 588--598},
\href{http://arxiv.org/abs/hep-ph/9607394}{{\ttfamily arXiv:hep-ph/9607394
  [hep-ph]}}.

\bibitem{Dimopoulos:1995mi}
S.~Dimopoulos and G.~Giudice, ``{Naturalness constraints in supersymmetric
  theories with nonuniversal soft terms},''
  \href{http://dx.doi.org/10.1016/0370-2693(95)00961-J}{{\em Phys.Lett.}
  {\bfseries B357} (1995) 573--578},
\href{http://arxiv.org/abs/hep-ph/9507282}{{\ttfamily arXiv:hep-ph/9507282
  [hep-ph]}}.

\bibitem{Sommer:1993ce}
R.~Sommer, ``{A New way to set the energy scale in lattice gauge theories and
  its applications to the static force and alpha-s in SU(2) Yang-Mills
  theory},'' \href{http://dx.doi.org/10.1016/0550-3213(94)90473-1}{{\em
  Nucl.Phys.} {\bfseries B411} (1994) 839--854},
\href{http://arxiv.org/abs/hep-lat/9310022}{{\ttfamily arXiv:hep-lat/9310022
  [hep-lat]}}.

\bibitem{Gockeler:2005rv}
M.~Gockeler, R.~Horsley, A.~Irving, D.~Pleiter, P.~Rakow, {\em et~al.}, ``{A
  Determination of the Lambda parameter from full lattice QCD},''
  \href{http://dx.doi.org/10.1103/PhysRevD.73.014513}{{\em Phys.Rev.}
  {\bfseries D73} (2006) 014513},
\href{http://arxiv.org/abs/hep-ph/0502212}{{\ttfamily arXiv:hep-ph/0502212
  [hep-ph]}}.

\bibitem{Juknevich:2009ji}
J.~E. Juknevich, D.~Melnikov, and M.~J. Strassler, ``{A Pure-Glue Hidden Valley
  I. States and Decays},''
  \href{http://dx.doi.org/10.1088/1126-6708/2009/07/055}{{\em JHEP} {\bfseries
  0907} (2009) 055},
\href{http://arxiv.org/abs/0903.0883}{{\ttfamily arXiv:0903.0883 [hep-ph]}}.

\bibitem{Juknevich:2009gg}
J.~E. Juknevich, ``{Pure-glue hidden valleys through the Higgs portal},''
  \href{http://dx.doi.org/10.1007/JHEP08(2010)121 10.1007/JHEP08(2010)121,
  10.1007/JHEP08(2010)121}{{\em JHEP} {\bfseries 1008} (2010) 121},
\href{http://arxiv.org/abs/0911.5616}{{\ttfamily arXiv:0911.5616 [hep-ph]}}.

\bibitem{Morningstar:1999rf}
C.~J. Morningstar and M.~J. Peardon, ``{The Glueball spectrum from an
  anisotropic lattice study},''
  \href{http://dx.doi.org/10.1103/PhysRevD.60.034509}{{\em Phys.Rev.}
  {\bfseries D60} (1999) 034509},
\href{http://arxiv.org/abs/hep-lat/9901004}{{\ttfamily arXiv:hep-lat/9901004
  [hep-lat]}}.

\bibitem{Chen:2005mg}
Y.~Chen, A.~Alexandru, S.~Dong, T.~Draper, I.~Horvath, {\em et~al.},
  ``{Glueball spectrum and matrix elements on anisotropic lattices},''
  \href{http://dx.doi.org/10.1103/PhysRevD.73.014516}{{\em Phys.Rev.}
  {\bfseries D73} (2006) 014516},
\href{http://arxiv.org/abs/hep-lat/0510074}{{\ttfamily arXiv:hep-lat/0510074
  [hep-lat]}}.

\bibitem{Bergner:2013nwa}
G.~Bergner, I.~Montvay, G.~M{\"u}nster, U.~D. {\"O}zugurel, and D.~Sandbrink,
  ``{Towards the spectrum of low-lying particles in supersymmetric Yang-Mills
  theory},'' \href{http://dx.doi.org/10.1007/JHEP11(2013)061}{{\em JHEP}
  {\bfseries 1311} (2013) 061},
\href{http://arxiv.org/abs/1304.2168}{{\ttfamily arXiv:1304.2168 [hep-lat]}}.

\bibitem{Bergner:2013jia}
G.~Bergner, I.~Montvay, G.~M{\"u}nster, D.~Sandbrink, and U.~D. {\"O}zugurel,
  ``{N=1 supersymmetric Yang-Mills theory on the lattice},'' {\em PoS}
  {\bfseries LATTICE2013} (2014) 483,
\href{http://arxiv.org/abs/1311.1681}{{\ttfamily arXiv:1311.1681 [hep-lat]}}.

\bibitem{Bergner:2014iea}
G.~Bergner, P.~Giudice, I.~Montvay, G.~M{\"u}nster, U.~D. {\"O}zugurel, {\em
  et~al.}, ``{Latest lattice results of N=1 supersymmetric Yang-Mills theory
  with some topological insights},'' {\em PoS} {\bfseries LATTICE2014} (2014)
  273,
\href{http://arxiv.org/abs/1411.1746}{{\ttfamily arXiv:1411.1746 [hep-lat]}}.

\bibitem{Bergner:2015iva}
G.~Bergner, P.~Giudice, G.~M{\"u}nster, S.~Piemonte, and D.~Sandbrink, ``{First
  studies of the phase diagram of N=1 supersymmetric Yang-Mills theory},'' {\em
  PoS} {\bfseries LATTICE2014} (2014) 262,
\href{http://arxiv.org/abs/1501.02746}{{\ttfamily arXiv:1501.02746 [hep-lat]}}.

\bibitem{Bergner:2015cqa}
G.~Bergner, P.~Giudice, G.~M{\"u}nster, and S.~Piemonte, ``{Witten index and
  phase diagram of compactified N=1 supersymmetric Yang-Mills theory on the
  lattice},''
\newblock 2015.
\newblock
\href{http://arxiv.org/abs/1510.05926}{{\ttfamily arXiv:1510.05926 [hep-lat]}}.
\newblock

\bibitem{Farrar:1997fn}
G.~R. Farrar, G.~Gabadadze, and M.~Schwetz, ``{On the effective action of N=1
  supersymmetric Yang-Mills theory},''
  \href{http://dx.doi.org/10.1103/PhysRevD.58.015009}{{\em Phys. Rev.}
  {\bfseries D58} (1998) 015009},
\href{http://arxiv.org/abs/hep-th/9711166}{{\ttfamily arXiv:hep-th/9711166
  [hep-th]}}.

\bibitem{Luty:1997fk}
M.~A. Luty, ``{Naive dimensional analysis and supersymmetry},''
  \href{http://dx.doi.org/10.1103/PhysRevD.57.1531}{{\em Phys. Rev.} {\bfseries
  D57} (1998) 1531--1538},
\href{http://arxiv.org/abs/hep-ph/9706235}{{\ttfamily arXiv:hep-ph/9706235
  [hep-ph]}}.

\bibitem{Evans:2013jna}
J.~A. Evans, Y.~Kats, D.~Shih, and M.~J. Strassler, ``{Toward Full LHC Coverage
  of Natural Supersymmetry},''
  \href{http://dx.doi.org/10.1007/JHEP07(2014)101}{{\em JHEP} {\bfseries 07}
  (2014) 101},
\href{http://arxiv.org/abs/1310.5758}{{\ttfamily arXiv:1310.5758 [hep-ph]}}.

\bibitem{Djouadi:2005gi}
A.~Djouadi, ``{The Anatomy of electro-weak symmetry breaking. I: The Higgs
  boson in the standard model},''
  \href{http://dx.doi.org/10.1016/j.physrep.2007.10.004}{{\em Phys. Rept.}
  {\bfseries 457} (2008) 1--216},
\href{http://arxiv.org/abs/hep-ph/0503172}{{\ttfamily arXiv:hep-ph/0503172
  [hep-ph]}}.

\bibitem{Djouadi:2005gj}
A.~Djouadi, ``{The Anatomy of electro-weak symmetry breaking. II. The Higgs
  bosons in the minimal supersymmetric model},''
  \href{http://dx.doi.org/10.1016/j.physrep.2007.10.005}{{\em Phys. Rept.}
  {\bfseries 459} (2008) 1--241},
\href{http://arxiv.org/abs/hep-ph/0503173}{{\ttfamily arXiv:hep-ph/0503173
  [hep-ph]}}.

\bibitem{Belanger:2013xza}
G.~Belanger, B.~Dumont, U.~Ellwanger, J.~F. Gunion, and S.~Kraml, ``{Global fit
  to Higgs signal strengths and couplings and implications for extended Higgs
  sectors},'' \href{http://dx.doi.org/10.1103/PhysRevD.88.075008}{{\em Phys.
  Rev.} {\bfseries D88} (2013) 075008},
\href{http://arxiv.org/abs/1306.2941}{{\ttfamily arXiv:1306.2941 [hep-ph]}}.

\bibitem{Djouadi:1997yw}
A.~Djouadi, J.~Kalinowski, and M.~Spira, ``{HDECAY: A Program for Higgs boson
  decays in the standard model and its supersymmetric extension},''
  \href{http://dx.doi.org/10.1016/S0010-4655(97)00123-9}{{\em Comput. Phys.
  Commun.} {\bfseries 108} (1998) 56--74},
\href{http://arxiv.org/abs/hep-ph/9704448}{{\ttfamily arXiv:hep-ph/9704448
  [hep-ph]}}.

\bibitem{Curtin:2013fra}
D.~Curtin, R.~Essig, S.~Gori, P.~Jaiswal, A.~Katz, T.~Liu, Z.~Liu, D.~McKeen,
  J.~Shelton, M.~Strassler, Z.~Surujon, B.~Tweedie, and Y.-M. Zhong, ``{Exotic
  decays of the 125 GeV Higgs boson},''
  \href{http://dx.doi.org/10.1103/PhysRevD.90.075004}{{\em Phys.Rev.}
  {\bfseries D90} no.~7, (2014) 075004},
\href{http://arxiv.org/abs/1312.4992}{{\ttfamily arXiv:1312.4992 [hep-ph]}}.

\bibitem{Curtin:2014pda}
D.~Curtin, R.~Essig, and Y.-M. Zhong, ``{Uncovering light scalars with exotic
  Higgs decays to $ b\overline{b}{\mu}^{+}{\mu}^{-} $},''
  \href{http://dx.doi.org/10.1007/JHEP06(2015)025}{{\em JHEP} {\bfseries 06}
  (2015) 025},
\href{http://arxiv.org/abs/1412.4779}{{\ttfamily arXiv:1412.4779 [hep-ph]}}.

\bibitem{Djouadi:2006bz}
A.~Djouadi, M.~M. Muhlleitner, and M.~Spira, ``{Decays of supersymmetric
  particles: The Program SUSY-HIT (SUspect-SdecaY-Hdecay-InTerface)},'' {\em
  Acta Phys. Polon.} {\bfseries B38} (2007) 635--644,
\href{http://arxiv.org/abs/hep-ph/0609292}{{\ttfamily arXiv:hep-ph/0609292
  [hep-ph]}}.

\bibitem{Beenakker:1996ed}
W.~Beenakker, R.~Hopker, and M.~Spira, ``{PROSPINO: A Program for the
  production of supersymmetric particles in next-to-leading order QCD},''
\href{http://arxiv.org/abs/hep-ph/9611232}{{\ttfamily arXiv:hep-ph/9611232
  [hep-ph]}}.

\bibitem{Sjostrand:2014zea}
T.~Sjöstrand, S.~Ask, J.~R. Christiansen, R.~Corke, N.~Desai, P.~Ilten,
  S.~Mrenna, S.~Prestel, C.~O. Rasmussen, and P.~Z. Skands, ``{An Introduction
  to PYTHIA 8.2},'' \href{http://dx.doi.org/10.1016/j.cpc.2015.01.024}{{\em
  Comput. Phys. Commun.} {\bfseries 191} (2015) 159--177},
\href{http://arxiv.org/abs/1410.3012}{{\ttfamily arXiv:1410.3012 [hep-ph]}}.

\bibitem{Drees:2013wra}
M.~Drees, H.~Dreiner, D.~Schmeier, J.~Tattersall, and J.~S. Kim, ``{CheckMATE:
  Confronting your Favourite New Physics Model with LHC Data},''
  \href{http://dx.doi.org/10.1016/j.cpc.2014.10.018}{{\em Comput. Phys.
  Commun.} {\bfseries 187} (2014) 227--265},
\href{http://arxiv.org/abs/1312.2591}{{\ttfamily arXiv:1312.2591 [hep-ph]}}.

\bibitem{deFavereau:2013fsa}
{\bfseries DELPHES 3} Collaboration, J.~de~Favereau, C.~Delaere, P.~Demin,
  A.~Giammanco, V.~Lema\^{i}tre, A.~Mertens, and M.~Selvaggi, ``{DELPHES 3, A
  modular framework for fast simulation of a generic collider experiment},''
  \href{http://dx.doi.org/10.1007/JHEP02(2014)057}{{\em JHEP} {\bfseries 02}
  (2014) 057},
\href{http://arxiv.org/abs/1307.6346}{{\ttfamily arXiv:1307.6346 [hep-ex]}}.

\bibitem{Cacciari:2011ma}
M.~Cacciari, G.~P. Salam, and G.~Soyez, ``{FastJet User Manual},''
  \href{http://dx.doi.org/10.1140/epjc/s10052-012-1896-2}{{\em Eur. Phys. J.}
  {\bfseries C72} (2012) 1896},
\href{http://arxiv.org/abs/1111.6097}{{\ttfamily arXiv:1111.6097 [hep-ph]}}.

\bibitem{Cacciari:2005hq}
M.~Cacciari and G.~P. Salam, ``{Dispelling the $N^{3}$ myth for the $k_t$
  jet-finder},'' \href{http://dx.doi.org/10.1016/j.physletb.2006.08.037}{{\em
  Phys. Lett.} {\bfseries B641} (2006) 57--61},
\href{http://arxiv.org/abs/hep-ph/0512210}{{\ttfamily arXiv:hep-ph/0512210
  [hep-ph]}}.

\bibitem{Cacciari:2008gp}
M.~Cacciari, G.~P. Salam, and G.~Soyez, ``{The Anti-k(t) jet clustering
  algorithm},'' \href{http://dx.doi.org/10.1088/1126-6708/2008/04/063}{{\em
  JHEP} {\bfseries 04} (2008) 063},
\href{http://arxiv.org/abs/0802.1189}{{\ttfamily arXiv:0802.1189 [hep-ph]}}.

\bibitem{Read:2002hq}
A.~L. Read, ``{Presentation of search results: The CL(s) technique},''
  \href{http://dx.doi.org/10.1088/0954-3899/28/10/313}{{\em J. Phys.}
  {\bfseries G28} (2002) 2693--2704}.
[,11(2002)].

\bibitem{TheATLAScollaboration:2013xia}
{\bfseries ATLAS} Collaboration, ``{Search for massive particles in multijet
  signatures with the ATLAS detector in $\sqrt{s} = 8$ TeV pp collisions at the
  LHC},'' {\em ATLAS-CONF-2013-091} (2013) .
\url{http://cds.cern.ch/record/1595753}.

\bibitem{Aad:2015lea}
{\bfseries ATLAS} Collaboration, G.~Aad {\em et~al.}, ``{Search for massive
  supersymmetric particles decaying to many jets using the ATLAS detector in
  $pp$ collisions at $\sqrt{s} = 8$ TeV},''
\href{http://arxiv.org/abs/1502.05686}{{\ttfamily arXiv:1502.05686 [hep-ex]}}.

\bibitem{Carloni:2010tw}
L.~Carloni and T.~Sjostrand, ``{Visible Effects of Invisible Hidden Valley
  Radiation},'' \href{http://dx.doi.org/10.1007/JHEP09(2010)105}{{\em JHEP}
  {\bfseries 09} (2010) 105},
\href{http://arxiv.org/abs/1006.2911}{{\ttfamily arXiv:1006.2911 [hep-ph]}}.

\bibitem{Carloni:2011kk}
L.~Carloni, J.~Rathsman, and T.~Sjostrand, ``{Discerning Secluded Sector gauge
  structures},'' \href{http://dx.doi.org/10.1007/JHEP04(2011)091}{{\em JHEP}
  {\bfseries 04} (2011) 091},
\href{http://arxiv.org/abs/1102.3795}{{\ttfamily arXiv:1102.3795 [hep-ph]}}.

\bibitem{ArkaniHamed:2001vr}
N.~Arkani-Hamed, A.~G. Cohen, and H.~Georgi, ``{Accelerated unification},''
\href{http://arxiv.org/abs/hep-th/0108089}{{\ttfamily arXiv:hep-th/0108089
  [hep-th]}}.

\bibitem{Craig:2011yk}
N.~Craig, D.~Green, and A.~Katz, ``{(De)Constructing a Natural and Flavorful
  Supersymmetric Standard Model},''
  \href{http://dx.doi.org/10.1007/JHEP07(2011)045}{{\em JHEP} {\bfseries 1107}
  (2011) 045},
\href{http://arxiv.org/abs/1103.3708}{{\ttfamily arXiv:1103.3708 [hep-ph]}}.

\end{thebibliography}\endgroup
\bibliographystyle{utphys}
\end{document}